\documentclass[pra,twocolumn,longbibliography]{revtex4-2}

\usepackage{braket}
\usepackage[T1]{fontenc}
\usepackage{amsmath}
\usepackage{amsthm}
\usepackage{amssymb}
\usepackage{amsfonts}
\usepackage{mathtools}
\usepackage{dsfont}
\usepackage[normalem]{ulem}
\usepackage{graphicx}
\usepackage{siunitx}
\usepackage{mhchem}
\usepackage{datetime}
\usepackage{enumitem}   
\usepackage{color}
\usepackage{newtxtext,newtxmath}
\usepackage{microtype}
\usepackage{braket}
\usepackage{xr}
\usepackage{natbib}
\usepackage{hyperref}
\usepackage{nicematrix}
\usepackage{multirow}
\usepackage[capitalise]{cleveref}
\hypersetup{%
    bookmarksopen=false,
    bookmarksnumbered=true,
    pdfnewwindow=true,
    unicode=false,
    colorlinks=true,%
    citecolor=blue,
    linkcolor=black,
    urlcolor=blue,
    filecolor=blue
    }
\theoremstyle{definition}

\newcommand{\partialder}[2]{\frac{\partial#1}{\partial#2}}

\renewcommand{\i}{\mathrm{i}}

\renewcommand{\Im}{\operatorname{Im}}

\renewcommand{\Re}{\operatorname{Re}}
\newcommand{\Tr}{\operatorname{Tr}}

\begin{document}
%%%%%%%%%%%%%%%%%%%%%%%%%%%%%%%%%%%%%%%%%%%%%%%%%%%%%%%%%%%%%%%%%%%%%%%%%%
%\newcommand{\figdir}{.}
\newcommand{\figdir}{figures}
%%%%%%%%%%%%%%%%%%%%%%%%%%%%%%%%%%%%%%%%%%%%%%%%%%%%%%%%%%%%%%%%%%%%%%%%%%
%%%affiliations
\newcommand{\liege}{Institut de Physique Nucléaire, Atomique et de Spectroscopie, CESAM, Universit\'e de Li\`ege, 4000 Liège, Belgium}

\newcommand{\oxford}{Department of Physics, University of Oxford, Clarendon Laboratory, Parks Road, Oxford OX1 3PU, United Kingdom}

\title{On-site interactions in quantum thermal machines: efficiency, rectification and entanglement beyond local and global master equations}

\newcommand{\comment}[1]{\textcolor{red}{#1}}

\author{
    Salvatore Araceli$^{1,\dagger}$,
    Teddy H. M. Ong$^{2,\dagger}$, 
    Baptiste Debecker$^{1}$,
    Kai Müller$^{3}$,
    Oliver Lunt$^{2}$,
    Andrew J. Daley$^{2}$,
    and François Damanet$^{1}$\\
    \small$^{1}$Institut de Physique Nucléaire, Atomique et de Spectroscopie, CESAM, Université de Liège, 4000 Liège, Belgium\\
    \small$^{2}$Department of Physics, University of Oxford, Oxford, United Kingdom\\
    \small$^{3}$Institut für Theoretische Physik, Technische Universität Dresden, D-01062 Dresden, Germany\\
    \small$^{\dagger}$These authors contributed equally to this work.
}
\begin{abstract}

Advances in experimental techniques have opened new routes for harnessing non-equilibrium dynamics in mesoscopic quantum systems. In this context, we study the impact of on-site interactions on the transport properties of a continuous quantum thermal machine composed of two coupled oscillators connected to two thermal reservoirs. In the weak system-reservoir coupling regime, where a long-standing debate concerns which reduced description should be preferred, we first show that the Redfield master equation (RME) provides an accurate and unifying framework that interpolates between two well-known limits: the \textit{local} and \textit{global} master equations. By relying on the Hierarchy of Pure States (HOPS), a numerically exact stochastic method, we then explore the full parameter space and show that interactions can be leveraged to tune the efficiency of the thermal machine at high temperatures (while leaving it essentially unchanged at low temperatures), induce non-reciprocal transport under asymmetric reservoir couplings, and generate steady-state entanglement within the junction. We derive expressions for system-bath correlators, such as heat and particle currents, consistently across different frameworks. Our work features on-site interactions to enhance the versatility of quantum thermodynamic junctions and clarifies the role of non-Markovianity and non-linearities in quantum transport.

\end{abstract}
\date{\today}
\maketitle

\section{Introduction}

Recent advances in designing and manipulating quantum systems have enabled experimental probes of quantum thermodynamics in a variety of platforms, including ultracold atomic gases \cite{bouton_quantum_2021,Koch2023Sep}, trapped ions \cite{maslennikov_quantum_2019}, nitrogen-vacancy centres \cite{Klatzow2019Mar}, superconducting qubits \cite{Aamir2025Thermally,Uusnakki2026May}, and quantum dots \cite{venturelli_minimal_2013}. In fact, it is possible to use just a single particle as the working medium \cite{von_lindenfels_spin_2019, rossnagelSingleatomHeatEngine2016, abah_single-ion_2012,Peterson2019Dec}. In turn, this has raised the possibility of turning to thermodynamic mechanisms for quantum information processing tasks. For example, efficient quantum refrigerators can reset a superconducting qubit by cooling it below the temperature of its direct surrounding environment~\cite{Aamir2025Thermally}.

Quantum thermal machines are designed to be either discrete, with work cycles analogous to classical thermodynamics~\cite{chen_interaction-driven_2019,rossnagelSingleatomHeatEngine2016,boettcher_dynamics_2024}, or continuous, by exchanging power with a time-dependent external drive~\cite{hoferMarkovianMasterEquations2017,leitch_driven_2022,bhandari_measurement-based_2023}. Although few-level quantum thermal machines of both types are well understood~\cite{revThermalengines,correa_quantum-enhanced_2014,kaur_effects_2025,linden_how_2010}, the role of interactions in more complex thermodynamic devices remains comparatively less explored. Interactions have been shown to reshape the performance of quantum heat engines, yielding, for instance, universal low-temperature regimes in many-particle working media~\cite{chen_interaction-driven_2019} or enhanced thermodynamic performance in few-body settings~\cite{Boubakour2023}. In related transport contexts, non-linearities can induce thermal rectification and negative differential thermal conductance in anharmonic junctions~\cite{Segal2006,Ruokola2009}, while numerically exact studies of strongly coupled anharmonic oscillators have highlighted the combined role of interactions and finite system-reservoir coupling in bosonic heat transport~\cite{Chen2020}. Yet, much less is known about continuous quantum thermal machines built from interacting systems with infinitely many levels and non-uniform spectra. In this regime, it remains unclear whether features such as efficiency at maximum power~\cite{lee_efficiency_2016,kaur_effects_2025,correa_quantum-enhanced_2014} remain universal, under which conditions entanglement can emerge alongside power production~\cite{huber_thermodynamic_2015,millen_perspective_2016,hovhannisyan_entanglement_2013}, and which reduced descriptions remain quantitatively reliable for predicting steady-state and transport properties.

Addressing this gap, we study an interacting, two-site oscillator junction between two thermal reservoirs, which provides a minimal setting for investigating both quantum thermal machines and quantum thermal transistors \cite{magazzu_thermal_2025, QuantumThermalTrans, poulsen_heat-based_2024}, whose behaviour depends on the directionality of external thermal bias. The addition of on-site interactions effectively interpolates between harmonic oscillators, anharmonic oscillators, and two-level systems. This both connects with a variety of relevant experimental set-ups, and allows for control over the transport and entanglement properties of the junction.

Describing the non-equilibrium dynamics of a junction coupled to several reservoirs is a challenging task. In the weak junction–reservoir coupling regime, a common strategy is to derive a master equation for the reduced density matrix of the junction. This simplifies the problem at the cost of approximations whose regimes of validity may be subtle. Much effort has therefore been devoted to comparing different master equation approaches, in particular the \textit{local} and \textit{global} descriptions~\cite{Levy2014, hoferMarkovianMasterEquations2017, Gonzales2017, Cattaneo_2019}. In the local approach, reservoirs act on individual subsystems of the junction, whereas in the global approach, the dissipation is constructed in the eigenbasis of the full junction Hamiltonian. 
These approaches are accurate in different parameter regimes, which motivates the search for a unified framework able to describe the crossover between them. Here, we show that the Redfield master equation (RME) provides such a framework: by retaining the relevant non-secular terms, it interpolates between both limits and enables a systematic study of transport across regimes, as partially anticipated in previous works~\cite{Scali2021localmaster, Potts2021}.

Beyond weak system-reservoir couplings, perturbative master-equation treatments are no longer expected to provide a controlled description, and more advanced approaches are required. For non-interacting, i.e., quadratic junctions, exact analytical or numerically exact results can be obtained using, for instance, non-equilibrium Green's function or scattering techniques~\cite{Wang2008, Dhar2008}. The situation is considerably more challenging for \textit{interacting} junctions strongly coupled to structured reservoirs. In this context, approaches based on Hierarchical Equations of Motion (HEOM) have been successfully used to study quantum heat transport and nonequilibrium charge transport beyond standard perturbative master equations~\cite{Kato2015,Kato2016,Schinabeck2016, Tanimura2020}. Here, we instead employ the Hierarchy of Pure States (HOPS)~\cite{suess_hierarchy_2014,hartmann_exact_2017,boettcher_dynamics_2024}, a stochastic pure-state formulation of non-Markovian open-system dynamics that corresponds to an unravelling of HEOM \cite{Suess2015Jun}. By propagating stochastic state vectors rather than the full reduced density matrix, HOPS reduces memory requirements in large Hilbert spaces, in close analogy with the advantage of Markovian quantum-trajectory methods over density-matrix master equations. Using HOPS, we show that the efficiency of the considered quantum thermal machine exhibits a strong robustness against on-site interactions at low temperatures. As a quantum transistor, we highlight the role of on-site interactions to control non-reciprocal transport and entanglement generation. Finally, we show that the mechanisms underlying our findings are directly reflected in momentum-resolved particle currents, which are measurable quantities and thus provide observable signatures of the interplay between non-Markovianity and interactions underpinning transport dynamics.

The rest of the manuscript is organised as follow. In Sec.~\ref{sec:model}, we define the two-site model and its coupling to the reservoirs. In Sec.~\ref{sec:me}, we derive the Redfield (RME), local (LME), and global (GME) master equations under different parameter regimes. In Sec.~\ref{sec:exact_hops}, we describe two numerical methods (HOPS and another method based on Heisenberg equations of motion for the closed system) that are used for benchmarking and reaching regimes outside the scope of master equations. We demonstrate that the RME encompasses the regimes of validity of both LME and GME. In Sec.~\ref{sec:transport}, we derive macroscopic thermodynamic quantities and microscopic transport quantities for RME and HOPS. In Sec.~\ref{sec:results}, we explore the system's behaviour and universal features. Parameter regimes required for non-reciprocity and entanglement generation are found. We further examine the effects of non-Markovianity by modifying the parameters of the bath spectral density. Finally, in Sec.~\ref{sec:conclusion}, we conclude and discuss a few possible extensions and applications of our work.

\section{Model}\label{sec:model}
\begin{figure}[t]
    \centering
\includegraphics[width=0.95\linewidth]{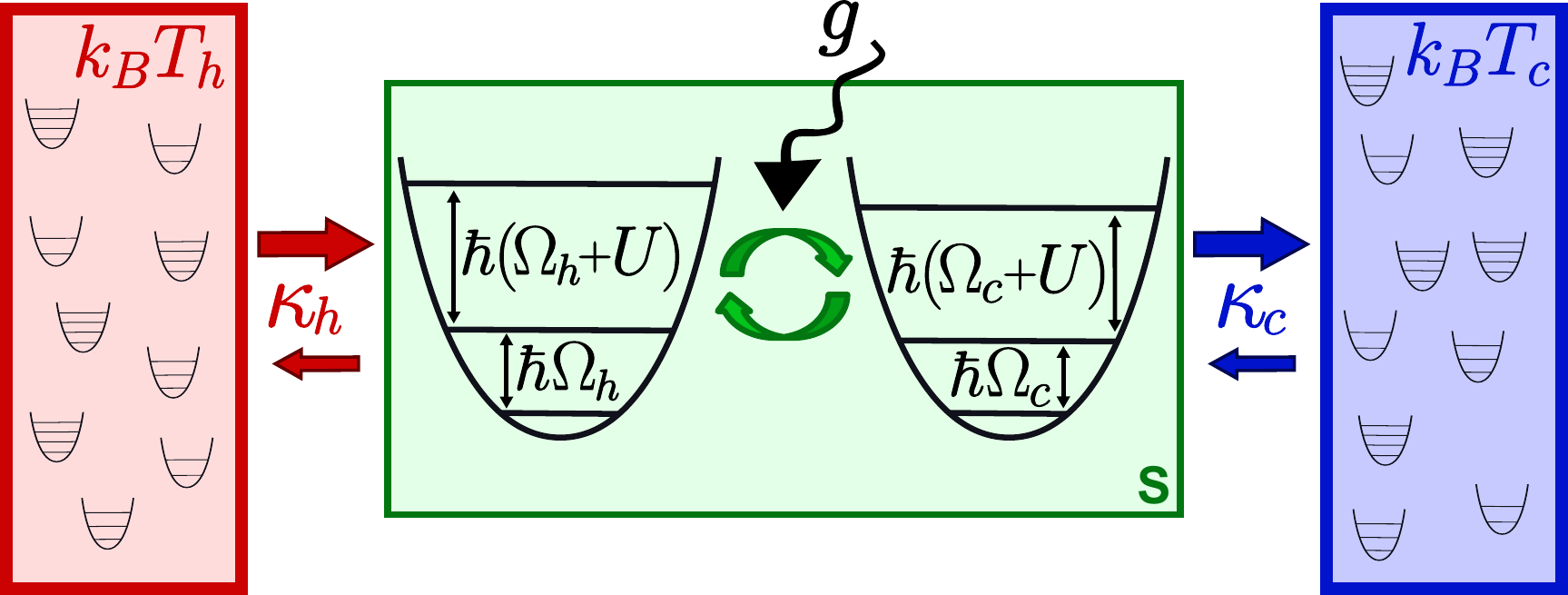}
    \caption{Sketch of the two-site junction of frequencies $\Omega_\ell$ ($\ell = h,c$) and on-site interactions $U$, coupled to two thermal reservoirs of temperature $T_\ell$. The tunnelling rate between the reservoirs and the junction is modulated by $\kappa_\ell$. The model also includes an external driving that couples the two sites with amplitude $g$. This setup can generate power from the thermal gradient between the two reservoirs, as heat currents flow through the junction.}
    \label{fig:1}
\end{figure}

We consider the quantum transport of bosonic particles through a two-site junction (with on-site interactions) connecting two reservoirs, as illustrated in Fig.~\ref{fig:1}. 

In the laboratory frame, the total Hamiltonian for the junction and the reservoirs is
\begin{equation}\label{eq:Hamiltonian_lab_1}
    H^\mathrm{lf}(t) = H_S^\mathrm{lf}(t) + H_h^\mathrm{lf} + H_c^\mathrm{lf}+H_I^\mathrm{lf},  
\end{equation}
where $H_S^\mathrm{lf}(t)$ is the junction Hamiltonian, $H^\mathrm{lf}_h$ and $H^\mathrm{lf}_c$ the hot (h) and cold (c) reservoir Hamiltonians, and $H_I^\mathrm{lf}$ the interaction Hamiltonian between the reservoirs and the junction. The superscript $^\mathrm{lf}$ denotes the laboratory frame. We assume that the Hamiltonian of the two-site junction takes the form
\begin{equation}\label{eq:Hamiltonian_lab_2}
   H_S^\mathrm{lf}(t) = \sum_\ell \left( \Omega_\ell a^\dagger_\ell a_\ell +\frac{U}{2} n_\ell (n_\ell-\mathds{1})\right) + g(a^\dagger_c a_h e^{\i\mathcal{E}t}+\mathrm{h.c.}),
\end{equation}
where $a_{\ell}$ ($\ell = h,c)$ is the annihilation operator of a bosonic particle of frequency $\Omega_\ell$ for the site $\ell$ of the junction (i.e., the site connected to the reservoir $\ell$), $g$ is the tunnelling rate between the two junction sites, $\mathcal{E} = \Omega_h -\Omega_c \geq 0$ is  the frequency of the external field, $n_\ell=a^\dagger_\ell a_\ell$ is the occupation of each site and $U\geq 0$ is the strength of the on-site interaction term, chosen to be identical for each site for simplicity. The junction reduces to free bosonic particles for $U=0$ and to qubits for $U\to\infty$, so that tuning $U$ makes it possible to interpolate between bosonic and spin statistics.

The reservoir Hamiltonians $H_h^\mathrm{lf}$ and $H_c^\mathrm{lf}$ read
\begin{equation}\label{eq:Hamiltonian_lab_3}
    H_\ell^\mathrm{lf} = \sum_k \omega_{\ell k} a_{\ell k}^\dagger a_{\ell k} \quad (\ell = h,c),
\end{equation}
where $a_{\ell k}$ ($\ell = h,c)$ is the annihilation operator of a bosonic particle of frequency $\omega_{\ell k}$ for the reservoir $\ell$. Finally, we assume a linear junction-reservoir Hamiltonian of the form
\begin{equation}\label{eq:Hamiltonian_lab_4}
    H_I^\mathrm{lf} = \sum_{\ell = h,c} \sum_k \left(g_{\ell k} a_\ell^\dagger a_{\ell k} + \mathrm{h.c.} \right),
\end{equation}
where $g_{\ell k}$ is the tunnelling rate between the junction site and the reservoir $\ell$. The rates $g_{\ell k}$ are chosen to be real and determine the bath spectral density $J_\ell(\omega)$, defined by 
\begin{equation}\label{def:bath_spectral_density}
    J_\ell(\omega) =  \sum_k g_{\ell k}^2 \delta(\omega-\omega_{\ell k}).
\end{equation}
In the following, we will focus on Ohmic spectral densities of the form
\begin{equation}
    J_\ell(\omega) = \frac{\kappa_\ell}{2 \pi \Omega_\ell}\omega e^{-\omega/\omega_\mathrm{cut}},
    \label{eq:ohmic}
\end{equation}
which have been investigated experimentally \cite{magazzu_probing_2018, peropadre_nonequilibrium_2013, forn-diaz_ultrastrong_2017}. A non-uniform spectral density enables investigation of the role of bath spectral structures on transport dynamics. Here, $\kappa_\ell$ is a parameter that tunes the magnitude of the system-bath coupling and $\omega_\mathrm{cut}$ denotes the cut-off frequency.

In the following, we choose to work in a rotating frame via the transformation
\begin{equation}
    {U^\mathrm{rf}}(t) = \mathrm{exp}\left[\mathrm{i}t\sum_\ell\Omega_\ell\left(a^\dagger_\ell a_\ell + \sum_k a^\dagger_{\ell k} a_{\ell k}\right)\right],
\end{equation}
which leads to the time-independent total Hamiltonian
\begin{equation}
\begin{aligned}\label{HamiltonianTotalRf}
    H &=  {U^\mathrm{rf}}(t)H^\mathrm{lf}(t) {U^\mathrm{rf}}^\dagger(t) -\mathrm{i}{U^\mathrm{rf}}(t)\partial_t {U^\mathrm{rf}}^\dagger(t) \\
    &= H_S + H_h + H_c + H_I,
\end{aligned}
\end{equation}
where
\begin{align}\label{Hamrotframe}
    &H_S = g(a_h^\dagger a_c + a_c^\dagger a_h)+\sum_\ell \frac{U}{2} n_\ell (n_\ell -\mathds{1}), \\
    &\label{Ham_Baths}H_\ell = \sum_k (\omega_{\ell k}-\Omega_\ell)a_{\ell k}^\dagger a_{\ell k},
\end{align}
and $H_I = H_I^\mathrm{lf}$ (as the interaction Hamiltonian is invariant under the rotating-frame). 

Note that the total Hamiltonian exhibits a $\mathbb{U}(1)$-symmetry associated with the operator $U_\mathrm{sym} = e^{\i\phi\left(a_h^\dagger a_h + a_c^\dagger a_c + \sum_{\ell, k}a_{lk}^\dagger a_{\ell k}\right)}$, i.e. for all $\phi \in \mathbb{R}$, we have $[H, U_\mathrm{sym} ] = 0$ and the total number of particles, $\langle a_h^\dagger a_h + a_c^\dagger a_c + \sum_{\ell, k}a_{\ell k}^\dagger a_{\ell k}\rangle$, is conserved.

\section{Master Equations}\label{sec:me}

In the weak junction-reservoir coupling regime, master equations describing the dynamics of the junction can be derived and solved analytically in some regimes, providing useful insights on the properties of the system. Here, we derive a Redfield master equation which, as anticipated in~\cite{purkayastha2017} and demonstrated below, provides a more accurate description of the junction properties and interpolates between the local and global approaches of~\cite{hoferMarkovianMasterEquations2017}, whose corresponding master equations are given below in the non-interacting case ($U = 0$) and in the qubit case ($U = \infty$) .

\subsection{Derivation}

The starting point is the usual nested Liouville Von-Neumann equation for the total density matrix in the interaction picture with respect to $H_S+H_h+H_c$, after tracing out the bath degrees of freedom. It reads
\begin{equation}
    \dot{\rho}^I = - \int_0^t  \mathrm{Tr}_B\left([H^I_I(t), [H_I^I(t-t'),\rho^I_{\mathrm{tot}}(t-t')] ]\right)dt'
\end{equation}
where $\rho^I = U^\dagger \rho U$, $\rho_{\mathrm{tot}}^I = U^\dagger \rho_\mathrm{tot}U$ and $H_I^I = U^\dagger H_I U $ are the junction and total density operators and the interaction Hamiltonian in the interaction picture, where $U(t) = e^{-\i (H_S+ H_h+H_c)t}$. Performing i) the Born approximation $\rho_{\mathrm{tot}}^I(t) \approx \rho^I(t)\otimes \rho_h^I \otimes \rho_c^I$ and ii) the Markov approximation, which consists in setting $\rho^I_{\mathrm{tot}}(t-t')\approx \rho^I_{\mathrm{tot}}(t)$ under the integral and pushing its upper limit to infinity, we obtain
\begin{equation}\label{EQ}
\begin{aligned}
&\dot{\rho}^I(t) = 
-  \sum_{\ell = h,c}\int_0^{\infty}dt' \\
\times&\bigg[ 
 \alpha_{\ell}(t -t') a_\ell^I(t) a_\ell^{I\dagger}(t') \rho^I(t)
+ \bar{\alpha}_{\ell}(t -t') a_\ell^{I\dagger}(t) a_\ell^I(t') \rho^I(t)\bigg. \\
&
- \bar{\alpha}_{\ell}^*(t -t') a_\ell^I(t) \rho^I(t)  a_\ell^{I\dagger}(t') 
- \alpha_{\ell}^*(t -t') a_\ell^{I\dagger}(t) \rho^I(t) a_\ell^I(t') \\[5pt] 
&
- \bar{\alpha}_{\ell}(t - t') a_\ell^I(t') \rho^I(t) a_\ell^{I\dagger}(t) 
- \alpha_{\ell}(t - t') a_\ell^{I\dagger}(t') \rho^I(t) a_\ell^I(t) \\
&\bigg.
+ \alpha_{\ell}^*(t - t') \rho^I(t) a_\ell^I(t') a_\ell^{I\dagger}(t) 
+ \bar{\alpha}_{\ell}^*(t - t') \rho^I(t) a_\ell^{I\dagger}(t') a_\ell^I(t) \bigg] 
\end{aligned}
\end{equation}
where $\alpha_{\ell}$ and $\bar{\alpha}_{\ell}$ are the bath correlation functions that are typically associated with the absorption and emission processes, respectively, and are defined as
\begin{align}\label{bcf1}
\alpha_{\ell}(s) &= \sum_k |g_{\ell k}|^2\langle a_{\ell k}^{I\dagger}(t) a_{\ell k}^I(0)\rangle \\
&=\int_0^{\infty} J_\ell(\omega)  n_\ell(\omega) e^{\i (\omega-\Omega_\ell) s} d\omega, \\
\bar{\alpha}_{\ell}(s) &= \sum_k |g_{\ell k}|^2\langle a_{\ell k}^{I}(t) a_{\ell k}^{I\dagger}(0)\rangle \\&= \int_0^{\infty} J_\ell(\omega) (1+n_\ell(\omega)) e^{- i (\omega - \Omega_\ell) s} d\omega.
\end{align}
Note that we assumed stationary reservoirs at thermal equilibrium at temperature $T_\ell$ ($\ell = h,c$), i.e., $\rho_\ell^I = e^{-H_\ell/T_\ell}/\mathrm{Tr_\ell[e^{-H_\ell/T_\ell}]}$. Performing the integration and coming back to the Schrödinger picture yields the Bloch-Redfield master equation (RME)
\begin{equation}\label{redfield}
\begin{aligned}
    \dot\rho &= - i [H_S, \rho] + \sum_{\ell = h,c}\left([L_{\ell} \rho, a_\ell] + [a_\ell^\dagger, \rho \,L_{\ell}^\dagger]\right) 
     \\
     &+ \sum_{\ell= h,c}\left([\bar{L}_\ell \rho, a_\ell^\dagger] + [a_\ell, \rho \,\bar{L}_{\ell}^\dagger]\right),
\end{aligned}
\end{equation}
with
\begin{align}
    L_\ell &= \int_0^\infty  \alpha_{\ell}(s) e^{-\i H_S s} a_\ell^\dagger\,  e^{\i H_S s} \mathrm{d}s,\label{def:L}
    \\
    \bar{L}_\ell &= \int_0^\infty \bar{\alpha}_{\ell}(s) e^{-\i H_S s} a_\ell\,  e^{\i H_S s} \mathrm{d}s.\label{def:Lbar}
\end{align}

Explicit expressions for \cref{def:L,def:Lbar} can be obtained by decomposing the operators $a_\ell$ and $a^\dagger_\ell$ into a sum of eigenoperators of the superoperator $[H_S,\cdot]$ \cite{breuerTheoryOpenQuantum2007}.

Below, we present the form of the master equation  in the two important limits $U = 0$ and $U = \infty$, which can be solved analytically.

\subsection{Non-interacting Case ($U = 0$)}

In the non-interacting regime ($U = 0$), the decomposition introduced above takes the explicit form
\begin{equation}
a_\ell = \frac{a_+ + \xi_\ell a_-}{\sqrt{2}},
\end{equation}
where $\xi_h = 1$ and $\xi_c = -1$. Here, the operators
\begin{equation}\label{defapm}
a_\pm = \frac{a_h\pm a_c}{\sqrt{2}}
\end{equation}
diagonalize the system Hamiltonian $H_S$ and satisfy the bosonic commutation relation $[a_\sigma, a^\dagger_{\sigma'}]= \delta_{\sigma\sigma'}$ for $\sigma,\sigma' \in \{+,-\}$. In this case, \cref{def:L,def:Lbar} reduce to
\begin{align}\label{def:L_for_U=0}
    L_{\ell} &= \frac{1}{\sqrt{2}}\left( \Gamma_{\ell}(\Omega_\ell +g) a_+^\dagger + \xi_\ell \Gamma_{\ell}(\Omega_\ell - g) a_-^\dagger  \right),
    \\ \label{def:Lbarre_for_U=0}
    \bar{L}_{\ell} &= \frac{1}{\sqrt{2}}\left( \bar{\Gamma}_{\ell}(\Omega_\ell +g) a_+ + \xi_\ell \bar{\Gamma}_{\ell}(\Omega_\ell - g) a_-  \right),
\end{align}
with the complex rates $\Gamma_{\ell}(\nu) = \gamma_{\ell}(\nu) + i \Delta_{\ell}(\nu)$ and $\bar{\Gamma}_{\ell}(\nu) = \bar{\gamma}_{\ell}(\nu) + i \bar{\Delta}_{\ell}(\nu)$ and the real rates $\gamma_\ell(\nu)$, $\bar{\gamma}_\ell(\nu)$, $\Delta_\ell(\nu)$ and $\bar{\Delta}_\ell(\nu)$ defined as
\begin{align}
    &\gamma_\ell(\nu) = \pi J_\ell(\nu)n_\ell(\nu),\nonumber \\
    &\bar{\gamma}_\ell(\nu) = \pi J_\ell(\nu)(n_\ell(\nu) +1),\nonumber\\
    &\Delta_\ell(\nu) = \int_0^\infty \frac{J_\ell(\omega) n_\ell(\omega)}{\omega - \nu} d\omega, \nonumber\\
    &\bar{\Delta}_\ell(\nu) = \int_0^\infty \frac{J_\ell(\omega) (n_\ell(\omega)+1)}{\nu - \omega} d\omega.
\end{align}

In this limit, the master equation can be cast into the standard form
\begin{equation}\label{Redfieldstd2}
\begin{split}
    \dot\rho& = \mathcal{L}^R[\rho] \\
    & = -\i[\tilde{H}_S^{R},\rho] +  \sum_{ij} L_{ij} (2C_i \rho C_j^\dagger - \{C_j^\dagger C_i,\rho\}),
    \end{split}
\end{equation}
with $C_i,C_j \in \{a_+,a_-,a_+^\dagger, a_-^\dagger\}$, $H_S^R= H_S +\sum_{ij} H^{R,LS}_{ij}C_i^\dagger C_j$, and the Kossakowski and Lamb-shifted Hamiltonian matrices $L$ and $H_S^R$ respectively given by
\begin{align}\label{Lmatrix}
    &L = \frac{1}{2}
    \begin{pmatrix}
        \bar{\gamma}_{h+} +\bar{\gamma}_{c+} & \bar{\Gamma} & 0 &0 \\
        \bar{\Gamma}^* &  \bar{\gamma}_{h-} +\bar{\gamma}_{c-} & 0 & 0 \\
        0 & 0 & \gamma_{h+}+ \gamma_{c+} & \Gamma \\
        0 & 0&  \Gamma^* & \gamma_{h-}+ \gamma_{c-}
    \end{pmatrix}, \\ \label{Hmatrix}
     &H_S^R =\frac{1}{2} \begin{pmatrix}
       \tilde{\Omega}_+& \beta & 0 & 0\\
       \beta^* &  \tilde{\Omega}_- & 0 & 0\\
       0 & 0 &  \tilde{\Omega}_+ & \beta^* \\
       0 & 0 & \beta &  \tilde{\Omega}_-
     \end{pmatrix},
\end{align}
where
\begin{align}
     &\Gamma = \sum_\ell \frac{\xi_\ell}{2}\left(\gamma_{\ell +} + \gamma_{\ell-}+ \i(\Delta_{\ell+}-\Delta_{\ell-})\right), \nonumber\\
     &\bar{\Gamma} = \sum_\ell \frac{\xi_\ell}{2} \left(\bar{\gamma}_{\ell +} + \bar{\gamma}_{\ell-}+ \i(\bar{\Delta}_{\ell+}-\bar{\Delta}_{\ell-})\right), \\
     &\tilde{\Omega}_\sigma = \xi_\sigma g + \frac{\sum_\ell(\bar{\Delta}_{\ell\sigma}+\Delta_{\ell\sigma})}{2}, \nonumber\\
    &\beta = \sum_{\substack{\ell = h,c \\\sigma = \pm}} \frac{\xi_\ell}{4}\left(\bar{\Delta}_{\ell\sigma}+\Delta_{\ell\sigma} + \i\xi_\sigma(\bar{\gamma}_{\ell\sigma}-\gamma_{\ell\sigma}) \right),
\end{align}
with $\xi_\pm = \pm 1$ and where we used the shortcuts
\begin{align}
      \gamma_{\ell \pm}\equiv \gamma_\ell(\Omega_\ell \pm g),&\quad \bar{\gamma}_{\ell \pm} \equiv \bar\gamma_\ell(\Omega_\ell \pm g),\nonumber  \\
    \Delta_{\ell\pm} \equiv \Delta_\ell(\Omega_\ell\pm g),&\quad \bar\Delta_{\ell\pm} \equiv \bar\Delta_\ell(\Omega_\ell\pm g).
\end{align}
Because the Kossakowski matrix $L$ is not necessarily positive semidefinite, Redfield master equations can generate unphysical states. However, we will work in regimes where this does not occur, i.e., where the Redfield map is completely positive. Furthermore, we will show that this Redfield description accurately captures the transport properties of the junction and interpolates between the local and global descriptions presented below in the case of $U = 0$, these latter relying on additional approximations with more restrictive regimes of validity. 

The \textit{local} approach makes the assumption that only the spectrum of the site directly coupled to the bath must be taken into account to write down the equations of motion \cite{hoferMarkovianMasterEquations2017,Cattaneo_2019}. This amounts to considering the approximation $a_\ell^I(t-s) \approx a_\ell^I(t)$
in Eq.~\eqref{EQ}, or taking the limit $g \to 0$ in the expression of $H_S$ appearing in the integrands of \cref{def:L,def:Lbar}. This results (in the Schrödinger picture) in the \textit{local} master equation (LME)
\begin{equation}\label{driving local}
\begin{aligned}
\dot \rho= \mathcal{L}^L[\rho] = &-i[\tilde{H}_S^L,\rho] +\sum_{\ell =h,c}\left(\gamma_\ell \mathcal{D}_{a_\ell^\dagger}
[\rho]+\bar{\gamma}_\ell\mathcal{D}_{a_\ell}[\rho]\right),
\end{aligned}
\end{equation}
where we define the dissipator as
\begin{equation}
   \mathcal{D}_{A}[\rho] =  2 A\rho A^\dagger - \{A^\dagger A,\rho\},
\end{equation}
and the renormalised Hamiltonian due to the Lamb shifts
\begin{equation}
    \tilde{H}_S^L = H_S + \sum_{\ell = h, c} (\bar\Delta_\ell + \Delta_\ell) a_\ell^\dagger a_\ell,
    \label{eq:Lamb}
\end{equation}
and where we use the shortcuts
\begin{align}
    \gamma_\ell \equiv \gamma_\ell(\Omega_\ell),&\quad
    \bar\gamma_\ell \equiv \bar\gamma_\ell(\Omega_\ell)\nonumber,\\ 
    \Delta_\ell \equiv \Delta_\ell(\Omega_\ell) ,&\quad \bar\Delta_\ell \equiv \bar\Delta_\ell(\Omega_\ell).
\end{align}

The \textit{global} approach consists of performing, on top of the Born and Markov approximations made before, the secular approximation in order to ensure the positivity of the resulting map~\cite{breuerTheoryOpenQuantum2007}. In contrast with the local approach, the global approach does take into account the full spectrum of the junction, but it neglects the so-called non-secular terms appearing in the master equation \cite{rivas_markovian_2010}. Those terms correspond to time-dependent terms in the interaction picture that oscillate as $e^{\pm 2 g i t}$ and involve pairs of $a_+$ or of $a_-$ operators.
Hence, removing those terms in Eq.~\eqref{redfield}, which corresponds to setting the off-diagonal terms of the matrices~\eqref{Lmatrix} and \eqref{Hmatrix} to zero, yields the \textit{global} master equation (GME)
\begin{equation}
\begin{aligned}\label{driving global}
\dot \rho(t) =&\mathcal{L}^G[\rho] = -\i[\tilde{H}_S^G,\rho(t)] \\ &+ \frac{1}{2}\sum_{\substack{\ell = h,c \\\sigma = \pm}}\left(\gamma_{\ell \sigma} \mathcal{D}_{a_\sigma^\dagger}[\rho(t)]+\bar{\gamma}_{\ell \sigma} \mathcal{D}_{a_\sigma}[\rho(t)]\right),
\end{aligned}
\end{equation}
with
\begin{equation}
    \tilde{H}_S^G = H_S + \frac{1}{2}\sum_{\substack{\ell = h,c \\\sigma = \pm}}(\bar{\Delta}_{\ell\pm} + \Delta_{\ell\pm})a_\sigma^\dagger a_\sigma.
\end{equation}

Because all the master equations above are quadratic, the dynamics of the system remains Gaussian and can be solved \textit{analytically}. App.~\ref{app:ME}  presents the analytical expressions for the steady state junction correlations for the three master equations derived in this section.

\subsection{Qubits ($U = \infty$)}

For $U = \infty$, the bosonic Hilbert spaces of each site of the junction can be reduced to the first two Fock states $|0\rangle_\ell$ and $|1\rangle_\ell$ so that they become effective two-level systems. The system Hamiltonian in the rotating frame now reads
\begin{equation}
    H_S = g(\sigma_+^h \sigma_-^c + \sigma_-^h \sigma_+^c),
\end{equation}
where $\sigma_-^\ell = |0\rangle \langle1|_\ell$ and $\sigma_+^\ell = |1\rangle \langle0|_\ell$.

The operators $L_\ell$  and $\bar{L}_\ell$ given by \cref{def:L,def:Lbar} are most conveniently  evaluated when working in the Hamiltonian eigenbasis.  The eigenvalues of $H_S$ are easily shown to be  $0$ (twice degenerate) and $\pm g$, while the corresponding eigenvectors read 
\begin{equation}
 \begin{split}
    &\ket{\phi_0^{(1)}} = \ket{0,0}, \quad  \ket{\phi_0^{(2)}} = \ket{1,1}, \quad \\ &\ket{\phi_{\pm g}} = \frac{\ket{0,1}\pm\ket{1,0}}{\sqrt{2}}.
    \end{split}
\end{equation}
Then, introducing the operators 
\begin{equation}
\tau_\ell =\frac{\tau_{+\ell}+\tau_{-\ell}}{\sqrt{2}},
\end{equation}
where $\tau_{+\ell} = A_1 + \xi_\ell A_2$, $\tau_{-\ell} = -\xi_\ell A_3 + A_4$ and
\begin{align}
    A_1 = \ket{\phi_0^{(1)}}\bra{\phi_{+g}}, &\quad A_2 = \ket{\phi_{-g}}\bra{\phi_0^{(2)}},\nonumber\\
     A_3 = \ket{\phi_0^{(1)}}\bra{\phi_{-g}}, &\quad A_4 = \ket{\phi_{+g}}\bra{\phi_0^{(2)}},
\end{align}
we obtain
\begin{align}\label{def:L_for_U=inf}
    L_{\ell} &= \frac{1}{\sqrt{2}}\left( \Gamma_{\ell}(\Omega_\ell +g) \tau_{+\ell}^\dagger + \Gamma_{\ell}(\Omega_\ell - g) \tau_{-\ell}^\dagger  \right),
    \\ \label{def:Lbarre_for_U=inf}
    \bar{L}_{\ell} &= \frac{1}{\sqrt{2}}\left( \bar{\Gamma}_{\ell}(\Omega_\ell +g) \tau_{+\ell} + \bar{\Gamma}_{\ell}(\Omega_\ell - g) \tau_{-\ell}  \right).
\end{align}
Note that the main difference between this case and the non interacting one lies in the fact that here, $\tau_{\pm h} \neq \tau_{\pm c}$, while they are the same for the first case. 
\section{Numerical Methods}\label{sec:exact_hops}

Beyond weak junction-reservoir coupling, the master equations derived in the previous section are not necessarily valid. To tackle stronger coupling, we rely on two numerically exact methods described below. For non-interacting particles $U=0$ and arbitrary junction-reservoir coupling, the evolution of the system and reservoirs can be simulated efficiently by exploiting the linearity of the Heisenberg equations of motion, referred to as the \textit{Exact} method (Sec.~\ref{subsec:exactmethod}). For interacting particles and arbitrary coupling, we use the Hierarchy Of Pure States (HOPS) (Sec.~\ref{sec:hops}).

To measure the closeness of two states obtained via different methods, we introduce the Uhlmann fidelity. Given the density matrices $\rho_1, \rho_2$, the fidelity between them is given by \cite{baldwin_efficiently_2023}
\begin{equation}\label{def:fidelity}
    \mathcal{F}(\rho_1,\rho_2)=\left[\mathrm{Tr}\left(\sqrt{\sqrt{\rho_{1}}  ~\rho_2~\sqrt{\rho_{1}}}\right)\right]^2.
\end{equation}

\subsection{Exact Method -- Heisenberg Equations of Motion ($U = 0$)}\label{subsec:exactmethod}

In order to compare the results obtained via the RME for non-interacting particles and study its validity in different regimes of parameters, we use the method introduced in \cite{rivas_markovian_2010}, which consists in considering the whole system-environment ensemble as a large closed system and in writing the Heisenberg equations of motion for all annihilation/creation operators from Eq.~(\ref{HamiltonianTotalRf}). The key ingredient here is that the total Hamiltonian is quadratic, making the equations linear. Thus, Gaussian states such as thermal states remain Gaussian through the whole evolution, which allows us to only compute the time evolution of the first and second moments in order to describe the dynamics of the whole system instead of all the elements of its density matrix. Implementation details are provided in App.~\ref{app:exact:formalism}, and the computation of fidelity between Gaussian states is explained in App.~\ref{app:exact:fidelity}.
\subsubsection{Comparison Between Exact Method and Master Equations}
\begin{figure}[t!]
    \centering
    \includegraphics[width=0.975\linewidth]{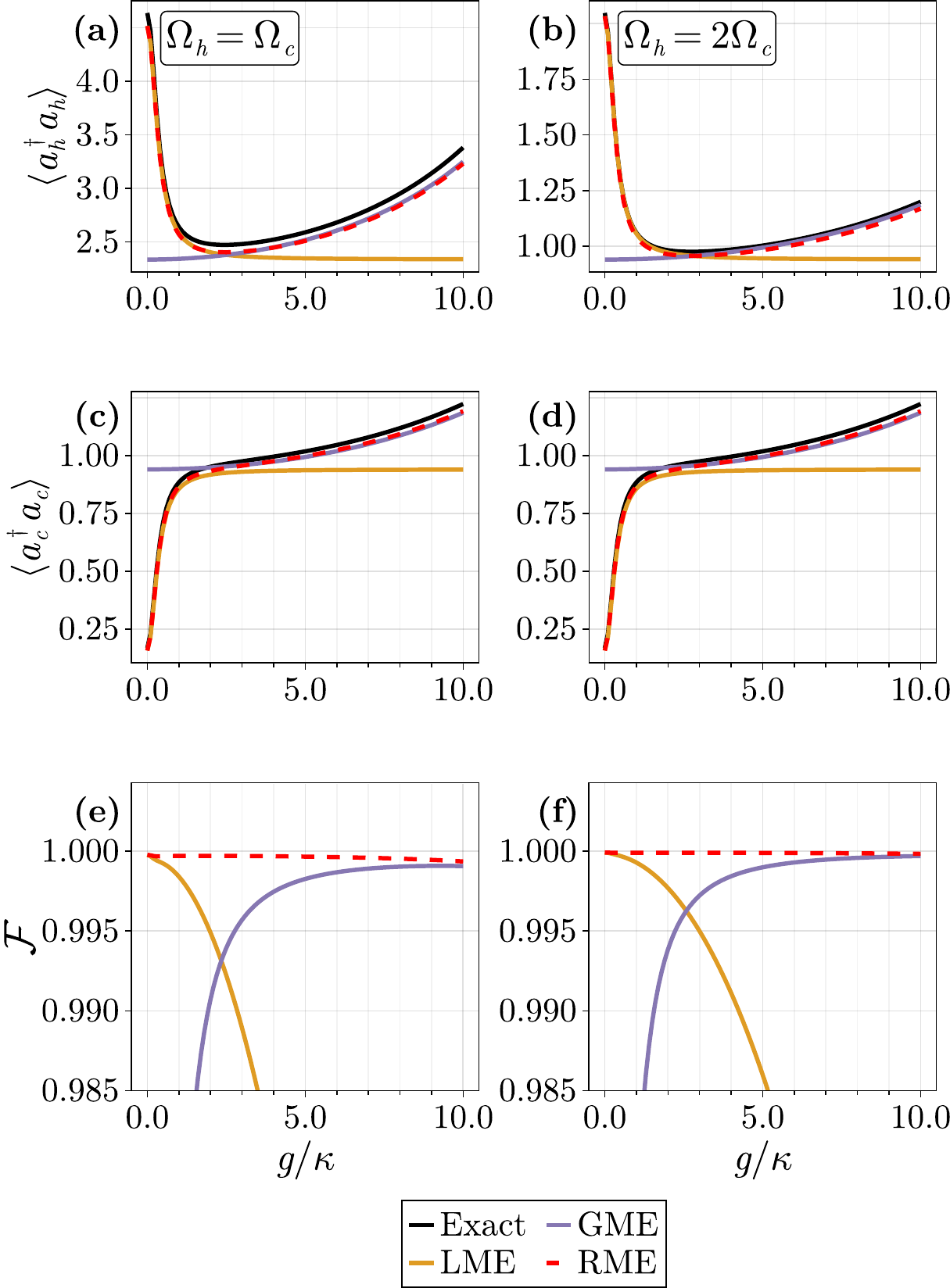}
    \caption{Comparison of steady-state properties obtained from the LME (Eq.~\eqref{driving local}), GME (Eq.~\eqref{driving global}), RME (Eq.~\eqref{Redfieldstd2}), and the Exact method for $U = 0$ as a function of the intercoupling strength $g/\kappa$.  Panels (a) and (b) (resp. (c) and (d)) show the populations of the left (resp. right) sites in two different setup: when the frequencies of the sites are taken the same (left column) and when they are not (right column). The RME results (red, dashed) interpolate perfectly between those of the LME (yellow), valid for small intercoupling, and the GME (purple), valid for large intercoupling. Panels (e) and (f) show the fidelity between the states obtained from the master equations and the state given by the exact method using \cref{def:fidelity}. We see that the RME outperforms the other two methods over the whole range of $g/\kappa$, resulting in a very high fidelity, above $0.999$.  Parameters: $ T_h/\Omega_c = 5.0, T_c/\Omega_c = 0.5, \kappa_{h,c}=\kappa=0.05\Omega_c, \omega_\mathrm{cut}/\Omega_c = 3.0, \Omega_c = 1.0, U=0$, panels (a), (c), and (e): $\Omega_h = \Omega_c$, panels (b), (d), and (f): $\Omega_h =2 \Omega_c$. Exact method parameters: $ N_\mathrm{bath} = 2000, t_\mathrm{evo} = 30/\kappa$.} 
    \label{fig:2}
\end{figure}

\begin{figure}[t]
    \centering
    \includegraphics[width=0.9\linewidth]{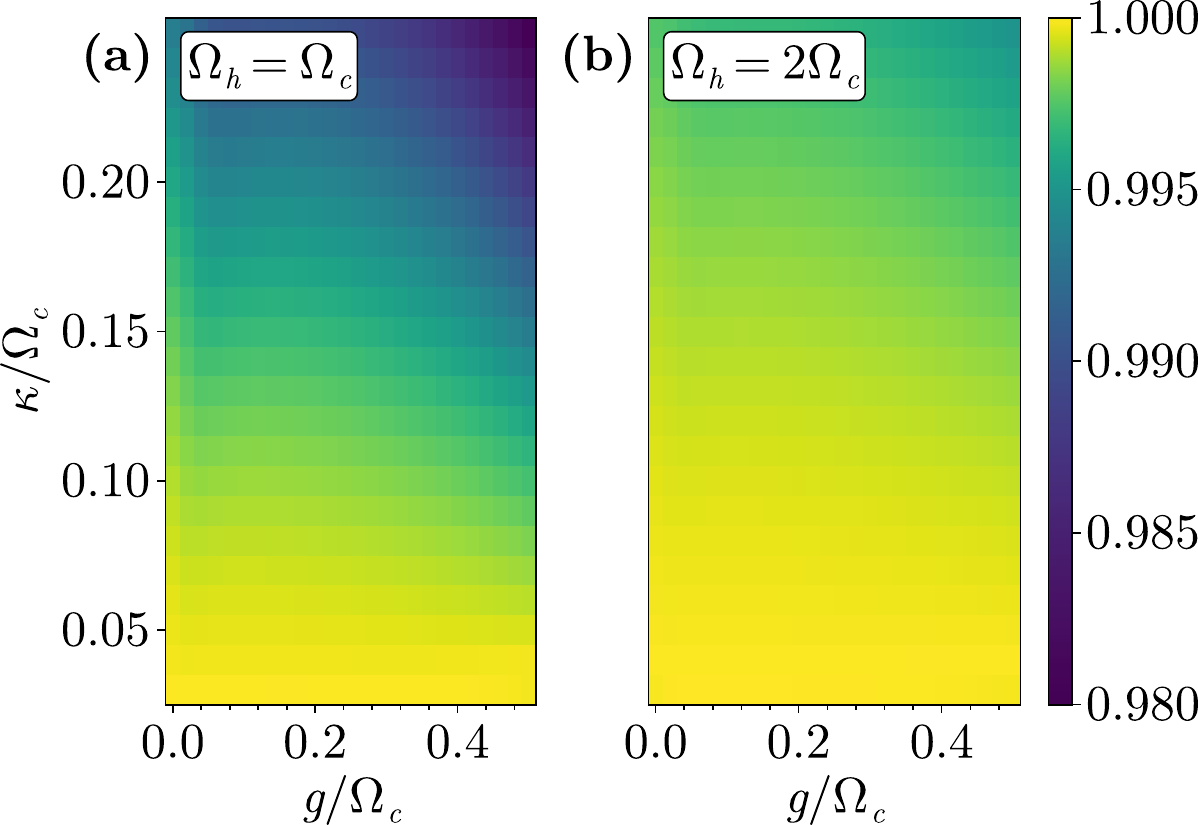}
    \caption{Fidelity (Eq.~\eqref{def:fidelity}) between the steady-states obtained by the RME and the Exact method as a function of $\kappa$ and $g$ for $\Omega_h = \Omega_c$ (panel (a)) and $\Omega_h = 2\Omega_c$ (panel (b)). As expected, the fidelity decreases for larger $\kappa$, but RME still predicts the state reasonably well even for $\kappa/\Omega_c = 0.25$, particularly when $\Omega_h> \Omega_c$. System parameters are the same as in Fig.~\ref{fig:2}. Exact method parameters: $ N_{\mathrm{bath}} = 2000, t_\mathrm{evo} = 30/\kappa$.}
    \label{fig:3}
\end{figure}

As a first important result, we show that, in the weak-coupling regime, the RME (\cref{Redfieldstd2}) provides accurate predictions for the steady states observables, interpolating between the LME (\cref{driving local}) and GME (\cref{driving global}) descriptions of the system known to have distinct regimes of validity, as displayed in Fig.~\ref{fig:2} for $U= 0$. This figure shows the steady-state populations of both junction sites as a function of $g/\kappa$ (with $\kappa_h = \kappa_c= \kappa$) in a similar fashion to \cite{hofer_markovian_2017} (panels (a)-(d)), as well as the fidelity between the steady states obtained from the three master equations considered and the one obtained from the exact global evolution (panels (e) and (f)) for two distinct cases: $\Omega_h = \Omega_c$ (left panels) and $\Omega_h = 2\Omega_c$ (right panels). Note that the first case corresponds to a non-driving situation, resulting in a time-independent Hamiltonian in the laboratory frame, already studied in \cite{rivas_markovian_2010}. For small $g/\kappa$, the predictions for both the LME and RME approaches are correct, though the fidelity for the LME drops quicker than for the RME. For large $g/\kappa$, the predictions for both the GME and RME are correct. Note that over the whole range of $g/\kappa$, the fidelity between the exact approach and the RME stays above 0.999.

The regime of validity of the RME can be extended beyond strict weak couplings to the baths. In Fig~\ref{fig:3}, we compare the steady state obtained from the RME and the Exact method. The fidelity between the two states is computed as a function of $g$ and $\kappa$ for $\Omega_h = \Omega_c$ (panel (a)) and $\Omega_h = 2\Omega_c$ (panel (b)), highlighting how valid the RME remains beyond small coupling. As expected, the fidelity decreases under strong system-reservoir coupling $\kappa$. However, even for $\kappa/\Omega_c = 0.25$, the RME still gives reasonable results, especially in the case $\Omega_h > \Omega_c$, which we interpret as a consequence of the resulting lower effective tunnelling between the two sites of the junction (due to the larger detuning), which makes the approximation of independent reservoirs more accurate.

\subsection{Hierarchy of Pure States (HOPS)}\label{sec:hops}

The Hierarchy of Pure States (HOPS) \cite{suess_hierarchy_2014, hartmann_exact_2017, hartmann_open_2021, boettcher_dynamics_2024} is an exact stochastic method for simulating non-Markovian open quantum systems. It can handle strong coupling regimes while placing no restrictions on the type of terms in the Hamiltonian of the system, including non-quadratic and time-dependent ones. In App.~\ref{app:HOPS:eom}, we describe the ingredients necessary to implement HOPS.

In this work, we modify the standard implementation in two ways (App.~\ref{app:HOPS:state}). Firstly, for $U=0$, a large number of Fock states are required for the harmonic oscillators at the temperatures considered. We reduce this by evolving the system under the action of displacement operators which diminish the occupation of higher Fock states.
Secondly, for $U>0$, the presence of the non-linear term $Un_\ell(n_\ell-1)/2$ makes the differential equation stiff and difficult to solve with conventional methods. To circumvent this, we place the system under the interaction picture of the non-linear term.
We discuss the formalism required to calculate observables in Apps.~ \ref{app:HOPS:obs}, \ref{app:HOPS:heat_current}, which yields formulae in the following sections. 

Arbitrary observables can be obtained by averaging correlations calculated from the state vector $\ket{\tilde{\psi}^{(\vec{0})}(t)}$ during individual trajectories. After a sufficiently long time, the expectation values fluctuate around their steady-state values, which we estimate by taking the mean over the converged section of the time series under the ergodic assumption. The error bars shown in all figures are twice the standard error of the mean, representing a $95\%$ confidence interval. Error bars for composite quantities, such as efficiency, are obtained using bootstrap resampling.

\section{Transport properties}\label{sec:transport}

In this section, we first define the fundamental quantities that we use to characterize the transport properties of the junction and how to compute them within the different frameworks.
Then, we describe how to compute momentum-resolved particle currents within our approaches.

\subsection{Power, Heat Currents and Efficiency}

In this work, we are mainly interested in the steady-state transport properties of the junction. 
The definitions of the power and the heat currents arise from the equation of motion of the total energy of the junction, i.e.,
\begin{equation}
\begin{split}
      \frac{\mathrm{d}\Tr [ H^\mathrm{lf}_S \rho^\mathrm{lf}(t)]}{\mathrm{d}t} & = \Tr\left[ \partialder{H^\mathrm{lf}_S}{t}\rho^\mathrm{lf}(t)\right]+ \i \Tr \left[ [H^\mathrm{lf},H^\mathrm{lf}_S] \rho^\mathrm{lf}(t)\right]\\
      &:= -\mathcal{P} +\sum_\ell  \mathcal{J}_\ell.
\end{split}
    \label{eq:def_currents}
\end{equation}
The first term corresponds to the extracted power $\mathcal{P}$ and the second term to the total heat current, which has contributions from both baths. This equation is none other than an interpretation of the first law of thermodynamics. At steady-state, the energy of the system is constant in time, resulting in $\mathcal{P} = \mathcal{J}_h+\mathcal{J}_c$. 

Using the lab frame Hamiltonian (\cref{eq:Hamiltonian_lab_1,eq:Hamiltonian_lab_2,eq:Hamiltonian_lab_3,eq:Hamiltonian_lab_4}), we find
\begin{align} 
        \mathcal{P} & =  -2 g\mathcal{E}\Im\left(\Tr[ a^\dagger_h a_c e^{-\i\mathcal{E}t} \rho^{\mathrm{lf}}(t)]\right), \label{def:power_exact_lab}\\
    \mathcal{J}_\ell & =  -2\sum_{ k}g_{\ell k}\Im\left(  \Tr[a_{\ell k}^\dagger S^{\mathrm{lf}}_\ell(t) \rho^{\mathrm{lf}}(t)] \right), \label{def:currents_exact_lab}
\end{align}
where
\begin{equation}
    S^{\mathrm{lf}}_\ell(t)  =  
    \begin{cases}
        \Omega_\ell a_\ell + ge^{\i\xi_{\bar{\ell}}\mathcal{E}t} a_{\bar{\ell}} +U n_\ell a_\ell & \text{if } U\geq 0,\\
       \Omega_\ell a_\ell + ge^{\i\xi_{\bar{\ell}}\mathcal{E}t} (\mathds{1}-2n_\ell)a_{\bar{\ell}} & \text{if } U\to \infty.
    \end{cases}
\end{equation}
Note that for the case of two qubits ($U\to\infty$), the commutation relations of truncated bosonic operators become $[a,a^\dagger]=\mathds{1}-2n$.
When transformed to the rotating frame (Eq.~\eqref{HamiltonianTotalRf}), Eqs.~\eqref{def:power_exact_lab}, \eqref{def:currents_exact_lab} become
\begin{align}
        \mathcal{P} & =  -2 g\mathcal{E}\Im\langle a^\dagger_h a_c \rangle,\label{def:power_rf}\\
    \mathcal{J}_\ell & =  -2\sum_{ k}g_{\ell k}\Im\left(\braket{a_{\ell k}^\dagger S_\ell^{\mathrm{rf}}} \right),\label{def:currents_exact_rf}
\end{align}
where
\begin{equation}
    S_\ell^{\mathrm{rf}}  = \begin{cases}
        \Omega_\ell a_\ell + g a_{\bar{\ell}} +U n_\ell a_\ell & \text{if } U\geq 0,\\
       \Omega_\ell a_\ell + g (\mathds{1}-2n_\ell)a_{\bar{\ell}} & \text{if } U\to \infty.
    \end{cases}
\end{equation}
The power and currents can be obtained readily from the Exact method (App.~\ref{app:exact:formalism}) when $U=0$.

In this work, we will define the efficiency of the thermodynamic device in the heat engine regime ($\mathcal{P},\mathcal{J}_h >0$),
\begin{equation}\label{def:efficiency}
    \eta := \frac{\mathcal{P}}{\mathcal{J}_h}.
\end{equation}

\subsubsection{Redfield Weak Coupling Regime}

In the weak-coupling regime, where the master equations are valid, the heat currents can be expressed in alternative forms involving junction correlations only. 

Labelling the different master equations we consider by $X = L,G,R$ (Local, Global, or Redfield), the heat current can be decomposed as
\begin{align}\label{currents}
   \sum_\ell &\mathcal{J}_\ell =  \mathrm{Tr}[H_S^\mathrm{lf}(t)\dot{\rho}^\mathrm{lf}(t)]\nonumber=\mathrm{Tr}[H_S^\mathrm{lf}(t) \mathcal{L}^{X,l}[\rho^\mathrm{lf}(t)]]\nonumber\\
   &=\mathrm{Tr}\left[H_S^\mathrm{lf}(t) \left(-\i[\tilde{H}_S^{X,\mathrm{lf}}(t),\rho^\mathrm{lf}(t)]+\mathcal{D}^{X,\mathrm{lf}}[\rho^\mathrm{lf}(t)] \right)\right].
\end{align}
For each master equation derived earlier in the rotating frame, we decomposed the effective Hamiltonian $\tilde{H}_S^X$ as $H_S + H_\mathrm{corr}^X$, which still holds in the lab frame. Thus, we have
\begin{align*}
    \mathrm{Tr}\left[H_S^\mathrm{lf} \left(-\i[\tilde{H}_S^{X,\mathrm{lf}},\rho^\mathrm{lf}] \right)\right]&= -\i \mathrm{Tr}\left[[H_S^\mathrm{lf},H_\mathrm{corr}^{X,\mathrm{lf}}]\rho^\mathrm{lf}\right]\\
    &=-\i\langle[\tilde{H}_S,H_\mathrm{corr}^{X,}]\rangle^X,
\end{align*}
where the last line is found by switching back to the rotating frame. The superscript $X$ for the correlations indicates that they should be evaluated using the solutions of the corresponding master equation and where \begin{equation}
    \tilde{H}_S =H_S+\sum_\ell \Omega_\ell a^\dagger_\ell a_\ell.
\end{equation} For the second term of Eq.~\eqref{currents}, we have $
    \mathcal{D}^{X,\mathrm{lf}}[\rho^\mathrm{lf}(t)] = {U^\mathrm{rf}}(t)^\dagger \mathcal{D}^{X}[\rho(t)] {U^\mathrm{rf}}(t)$ and thus
\begin{equation*}
    \mathrm{Tr}\left[H_S^\mathrm{lf}(t) \mathcal{D}^{X,\mathrm{lf}}[\rho^\mathrm{lf}(t)]\right] = \mathrm{Tr}\left[\tilde{H}_S \mathcal{D}^{X}[\rho(t)]\right].
\end{equation*}
Finally, for each master equation, we have the following decompositions, $
    H^X_\mathrm{corr} = \sum_\ell H^X_{\mathrm{corr},\ell}$ and $
    \mathcal{D}^X[\rho(t)] = \sum_\ell \mathcal{D}_\ell^X[\rho(t)]$, i.e. the total dissipation can be interpreted as a sum of dissipative terms coming from the hot and cold reservoirs. Thus,  we find
\begin{equation}\label{Current l for ME}
    \mathcal{J}_\ell^X = -\i\langle[\tilde{H}_S,H_{\mathrm{corr},\ell}^{X,}]\rangle^X + \mathrm{Tr}\left[\tilde{H}_S \mathcal{D}_\ell^{X}[\rho(t)]\right].
\end{equation}
It is important to note that, in the Redfield and Local cases ($X = R, L$)  the first term of Eq.~\eqref{Current l for ME} does not vanish because $H_{\mathrm{corr}}$ does not commute with $\tilde{H}_S$. This is an important distinction between the Redfield and the local approaches with the global one. 

Given this, in terms of junction correlations, the heat currents for Redfield read
\begin{align}\label{def:heat_current_Redfield}
    \mathcal{J}_{\ell}^R &= \sum_{\sigma=\pm} \left( \pi J_{\ell\sigma} \big[ \Omega_{\ell\sigma}(n^{\ell\sigma}-\langle a^\dagger_\sigma a_\sigma\rangle^R)\right.\nonumber\\
    &\left. \quad- \xi_\ell \Omega_{\ell\bar{\sigma}}\Re[\langle a^\dagger_+ a_-\rangle^R] \big] \right. \nonumber \\
    &\quad \left. + \xi_\ell\xi_\sigma\Im[\langle a^\dagger_+ a_-\rangle^R]\Omega_{\ell\sigma}(\Delta_{\ell\bar{\sigma}}+\bar{\Delta}_{\ell\bar{\sigma}}) \right)
\end{align}

For the local approach, we have
\begin{align}\label{powerlocalSS}
\mathcal{J}^L_\ell =&2 g (\bar{\Delta}_\ell+\Delta_\ell)\xi_\ell\Im[\langle a^\dagger_h a_c\rangle^L]\nonumber\\
+&2\pi J_\ell\left[\Omega_\ell(n^\ell-\langle a_\ell^\dagger a_\ell\rangle^L) -g\Re[\langle a_h^\dagger a_c\rangle^L] \right]. 
\end{align}

For the global approach, we must change the definition of the power and the heat currents because the bath couples to the eigenmodes of the system Hamiltonian, which depend directly on the external driving $g$. Consistent with the definition introduced in \cite{revThermalengines,hoferMarkovianMasterEquations2017,QuantRefThirdLaw,GELBWASERKLIMOVSKY2015329}, we define the total current associated with a given bath as the sum of the individual mode currents corresponding to that bath.  With this, we can define the heat current in the steady-state for the bath $\ell$ as
\begin{equation}\label{Def:current_global}
\begin{aligned}
    \mathcal{J}^G_{\ell} &= -\frac{1}{\beta_\ell}\sum_\sigma\Tr\{ \mathcal{L}_{\ell \sigma}[\rho_\mathrm{SS}]\ln \rho_{\ell \sigma}\}\\
    &= \sum_\sigma \Omega_{\ell\sigma}\pi J_{\ell\sigma}\left( n^{\ell\sigma}-\langle a^\dagger_\sigma a_\sigma\rangle^G\right).
\end{aligned}
\end{equation}
This expression correspond exactly to the one given in \cref{def:heat_current_Redfield} where instead of evaluating the correlations with the RME solutions, we use the GME's ones.
Note that in the global approach, the definition of the power comes from enforcing the first law, i.e. $\mathcal{P}^G_\mathrm{SS} = \mathcal{J}^G_{h}+\mathcal{J}^G_{c}$. Indeed, as $\Im\braket{a^\dagger_h a_c}^G = 0$, the definition of \cref{def:power_rf} would predict no power for every regime of parameters.

For the qubit case ($U\to\infty$), we obtain the heat currents through \cref{currents} taken in the rotating frame and using the corresponding Liouvillian.

\subsubsection{HOPS}

Consider the trajectory-wise evolution of the total energy of the system $\mathbb{E}\left[H^\mathrm{hf}_S\right]$,
where $\mathbb{E}$ denotes taking the expectation value using the normalised state vector $\ket{\tilde{\psi}^{(\vec{0})}(t)}$, and $\mathrm{hf}$ denotes the rotating frame used by HOPS (App.~\ref{app:HOPS:obs}).  The equation of motion is as follows, 
\begin{equation}
\begin{split}
\frac{\mathrm{d}~\mathbb{E}\left[H^\mathrm{hf}_S\right]}{\mathrm{d}t} &= \mathbb{E}\left[\partialder{H^\mathrm{hf}_S}{t}\right]+ \bra{\tilde{\psi}^{(\vec{0})}(t)}H^\mathrm{hf}_S \frac{d\ket{\tilde{\psi}^{(\vec{0})}(t)}}{dt} +\mathrm{h.c.}.
\end{split}
    \label{eq:def_currents_HOPS}
\end{equation}
We define the HOPS empirical estimate for power and heat currents as 
\begin{align}
        \mathcal{P}&=-\mathcal{M}\left\{\mathbb{E}\left[\partialder{H^\mathrm{hf}_S}{t}\right]\right\},\label{def:HOPS:power}\\
        \sum_\ell \mathcal{J}_\ell &= \mathcal{M}\left\{\bra{\tilde{\psi}^{(\vec{0})}(t)}H^\mathrm{hf}_S \frac{\mathrm{d}\ket{\tilde{\psi}^{(\vec{0})}(t)}}{\mathrm{d}t} +\mathrm{h.c.}\right\},\label{def:HOPS:currents}
\end{align}
where $\mathcal{M}$ denotes the mean over HOPS trajectories. In App.~\ref{app:HOPS:heat_current}, we detail how these quantities are computed to fourth-order time-step accuracy under the RK4 scheme. 
%As the HOPS equation of motion is non-linear, the results are not guaranteed to be linear in the system's reduced density matrix $\rho(t)=\mathcal{M}\left\{\ket{\tilde{\psi}^{(\vec{0})}(t)}\bra{\tilde{\psi}^{(\vec{0})}(t)}\right\}$, in contrast with predictions by the master equations.

We verify the validity of the empirical expressions (\cref{def:HOPS:power,def:HOPS:currents}) by using the HOPS formalism (App.~\ref{app:HOPS:obs}) to directly evaluate Eq.~\eqref{def:currents_exact_lab} in terms of accessible quantities, obtaining excellent agreement in their steady-state values (App.~\ref{app:HOPS:comparison}). The data presented in this paper are calculated using the empirical expression, as it presents more flexibility in general use cases.

\subsection{Momentum-resolved particle currents} 
To obtain useful physical insights on the dynamics of the system, it is also worthwhile to track the momentum-dependent particle currents in the reservoirs. For a Markovian bath weakly coupled to a system, this would be a sum of Lorentzians whose widths are proportional to the coupling strength and whose centres are displaced from the system's eigenfrequencies due to Lamb shifts~\cite{jin_generic_2020}. Non-Markovianity, in the form of a non-trivial spectral density, can weight different parts of the spectrum, while non-linearity in the system can mix eigenfrequencies.

\subsubsection{Exact Method}
From the Heisenberg equation of motion, bath modes observables are accessible, including $ \langle a_{k\ell}^\dagger a_{k\ell}\rangle$. We obtain $  \dot{\langle a_{k\ell}^\dagger a_{k\ell}\rangle}$ by performing Ordinary Least Squares (OLS) regression \cite{rao_linear_1973} on the time-series data of the occupation $  \langle a_{k\ell}^\dagger a_{k\ell}\rangle(t)$.

\subsubsection{Redfield Weak coupling regime}
Note that the dynamics of the bath correlations can be related to the correlation functions of the junction. Indeed, because 
\begin{equation}
   \dot{a}_{k\ell}  = - i (\omega_{k\ell} - \Omega_\ell) a_{k\ell} - i g_{k\ell} a_\ell,
\end{equation}
we can easily write the momentum-resolved particle current into the reservoir as
\begin{multline}
    \dot{\langle a_{k\ell}^\dagger a_{k\ell}\rangle} \\= 2 \Re\left\{ |g_{k\ell}|^2\ \int_{0}^t e^{-\i (\omega_{k\ell} - \Omega_\ell)(t-t')} \langle a_\ell^\dagger(t) a_{\ell}(t')\rangle dt' \right\},
\end{multline}
where we used the Born approximation and the fact that the initial reservoir state is thermal. In the steady state, the junction correlation function inside the integral depends only on the time difference, and we can write
\begin{equation}
\begin{aligned}
    &\dot{\langle a_{k\ell}^\dagger a_{k\ell}\rangle} = 2 \,\Re\left\{ |g_{k\ell}|^2\ \int_{0}^\infty e^{-\i (\omega_{k\ell} - \Omega_\ell)\tau} \langle a_\ell^\dagger(\tau) a_{\ell}(0)\rangle_{ss} d\tau \right\} \\ &= 2 \,\Re\left\{ |g_{k\ell}|^2\ \mathrm{Tr}[a_\ell^\dagger\left(\int_{0}^\infty e^{-\i (\omega_{k\ell} - \Omega_\ell)\tau + \mathcal{L}\tau}\right)  a_{\ell} \rho_{ss}] d\tau  \right\} \\
    \\&= 2 \,\Re\left\{ |g_{k\ell}|^2\ \mathrm{Tr}[a_\ell^\dagger \big(i(\omega_{k\ell} - \Omega_\ell) \mathds{1} -\mathcal{L}\big)^{-1}  [a_{\ell}\rho_{ss}] ]\right\},\\
\end{aligned}
\end{equation}
where $\mathcal{L}$ is the Liouvillian of the Local, Global, or Redfield approaches. Hence, from the knowledge of the steady state and of the Liouvillian, one can calculate the momentum-resolved particle current into the reservoir. Finally, we define the normalised particle current
\begin{equation}
    \mathcal{Q}_\ell(\omega):= \frac{
    1}{\Delta \omega} \dot{\langle a_{k\ell}^\dagger a_{k\ell}\rangle} ~\delta(\omega - \omega_{\ell k}),
    \label{def:particle_current}
\end{equation}
where $\Delta \omega=\omega_{\ell ~k+1}-\omega_{\ell k}$ is the uniform discretisation chosen for the bath modes.
\begin{figure*}[ht]
    \centering
\includegraphics[width=\linewidth]{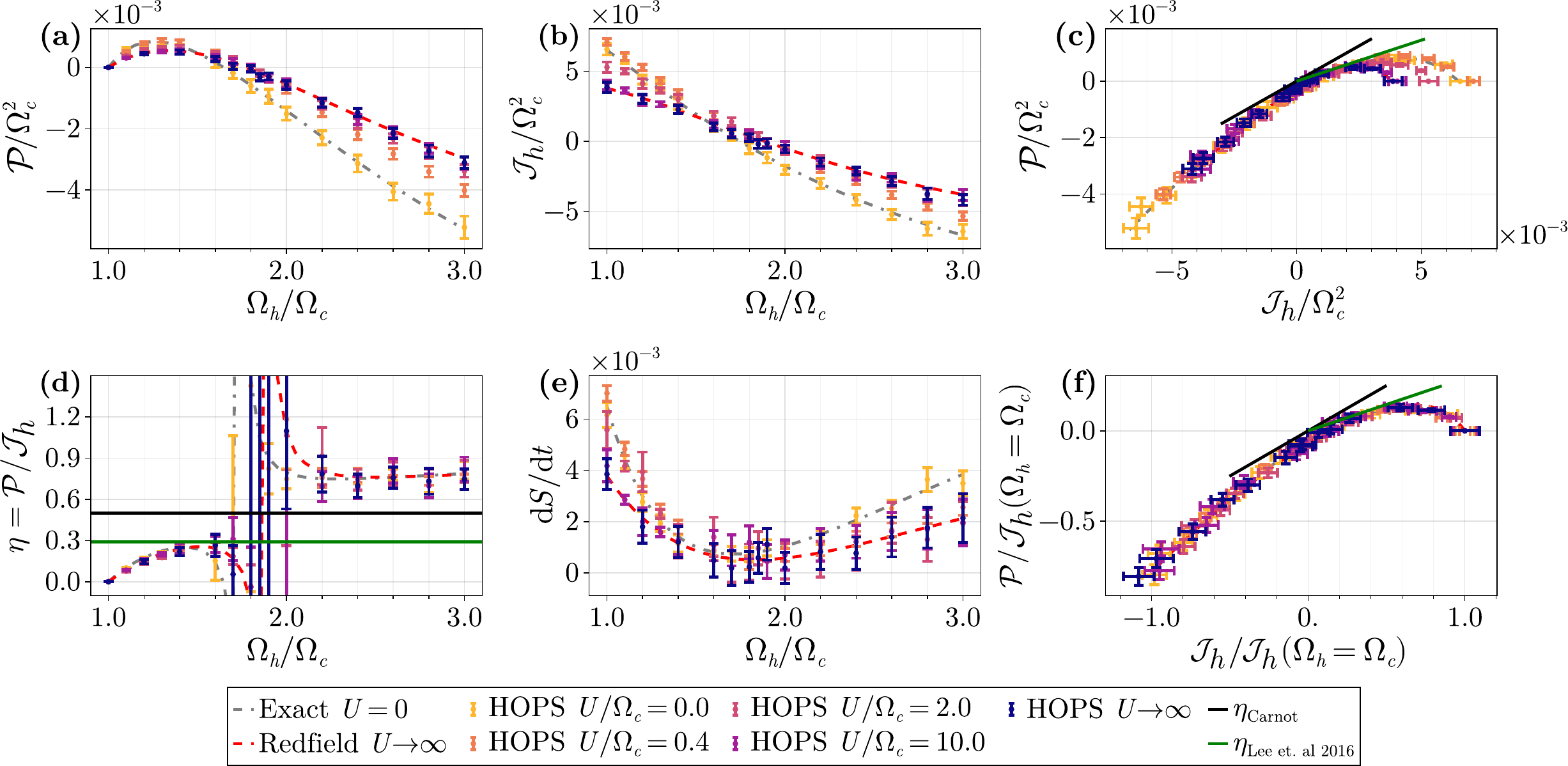}
    \caption{Thermodynamic quantities of the thermal machine at low temperatures.
    (a) Power $\mathcal{P}$ (Eq.~\eqref{def:HOPS:power} for HOPS, \cref{def:power_rf} for Redfield and Exact method) is positive in the heat engine regime and negative in refrigerator regime. The stalling point $\lambda_{\mathrm{stall}}$ increases with $U$. The value of $\Omega_h/\Omega_c$ where maximum power is produced also increases.
    (b) Heat current from hot bath $\mathcal{J}_h$ (Eq.~\eqref{def:HOPS:currents} for HOPS, Eq.~\eqref{def:currents_exact_rf} for Redfield and Exact method) decreases monotonically with increasing $\Omega_h/\Omega_c$.
    (c) Power $\mathcal{P}$ against heat current $\mathcal{J}_h$ shows that increasing $U$ decreases the magnitudes of thermodynamic quantities. (d) Efficiency $\eta$ (Eq.~\eqref{def:efficiency}) increases with $\Omega_h/\Omega_c$ in the heat engine regime, but diverges near $\lambda_{\mathrm{stall}}$. Observe that the efficiencies for different $U$ are very similar. (e) Entropy production rate $\mathrm{d}S/\mathrm{d}t$ (Eq.~\eqref{def:entropy}) is always positive, in accordance with the Second Law of Thermodynamics. (f) Power $\mathcal{P}$ against heat current $\mathcal{J}_h$ rescaled by $\mathcal{J}_h(\Omega_h=\Omega_c)$, resulting in the collapse of the curves from (c) and demonstrating `universal' behaviour. The lines corresponding to the Carnot efficiency and that predicted by \cite{lee_efficiency_2016} are illustrated as straight lines in panels (c), (d), (f).
    System parameters: $T_h/\Omega_c=1.0, T_c/\Omega_c=0.5, \kappa_{h,c}=0.05\Omega_c, \omega_{\mathrm{cut}}/\Omega_c=3.0, \Omega_h/\Omega_c\in[1.0, 3.0], g/\Omega_c = 0.5$. HOPS parameters are listed in Table~\ref{table:HOPS:params}. Exact method parameters: $N_{\mathrm{bath}}=2000, t_{\mathrm{evo}}=30/\kappa_c$.} 
    \label{fig:U:thermo_T_10}
\end{figure*}

\subsubsection{HOPS}
The occupancy $\braket{a_{\ell k}^\dagger a_{\ell k}}$ of the frequency mode $\omega_{\ell k}$ in bath $\ell=h,c$ is given by
\begin{equation}
    \begin{split}
        \braket{a_{\ell k}^\dagger a_{\ell k}}(t) & = \mathcal{M} \left\{ \left(|\tilde{z}_{\ell k}(t) + y_{\ell k} |^2 -1\right)\right\},
    \end{split}
\end{equation}
\noindent
where $\tilde{z}_{\ell k}(t)$ and $y_{\ell k}$ are elements that form coloured noises $\tilde{\eta}_\ell(t), \zeta_\ell(t)$ (derivation given in App.~\ref{app:HOPS:bath_occupancy}). Matching the calculation done for the Exact method, we compute the current $\dot{\langle a_{k\ell}^\dagger a_{k\ell}\rangle}$ by performing an OLS regression on the time-series data at the trajectory level (App.~\ref{app:HOPS:gradient}).

\section{Results}\label{sec:results}
We now turn to examine the impact of the repulsive interactions $U$ on the transport properties of the junction. We first consider the effects of $U$ on its thermodynamic properties. Then, we study how the interactions can transform the junction into a rectifier, and analyse the entanglement properties of the two-site system. In order to provide a better understanding of the underlying mechanisms yielding the observed effects, we also study the influence of the form of the bath spectral density and its interplay with the system on-site interactions. Finally, we discuss the behaviour of the momentum-resolved bath transport spectrum, a measurable quantity, making it possible to highlight which signatures of our results could be observed in current experiments.

\subsection{Thermodynamic properties}
\begin{figure*}[ht]
    \centering
\includegraphics[width=\linewidth]{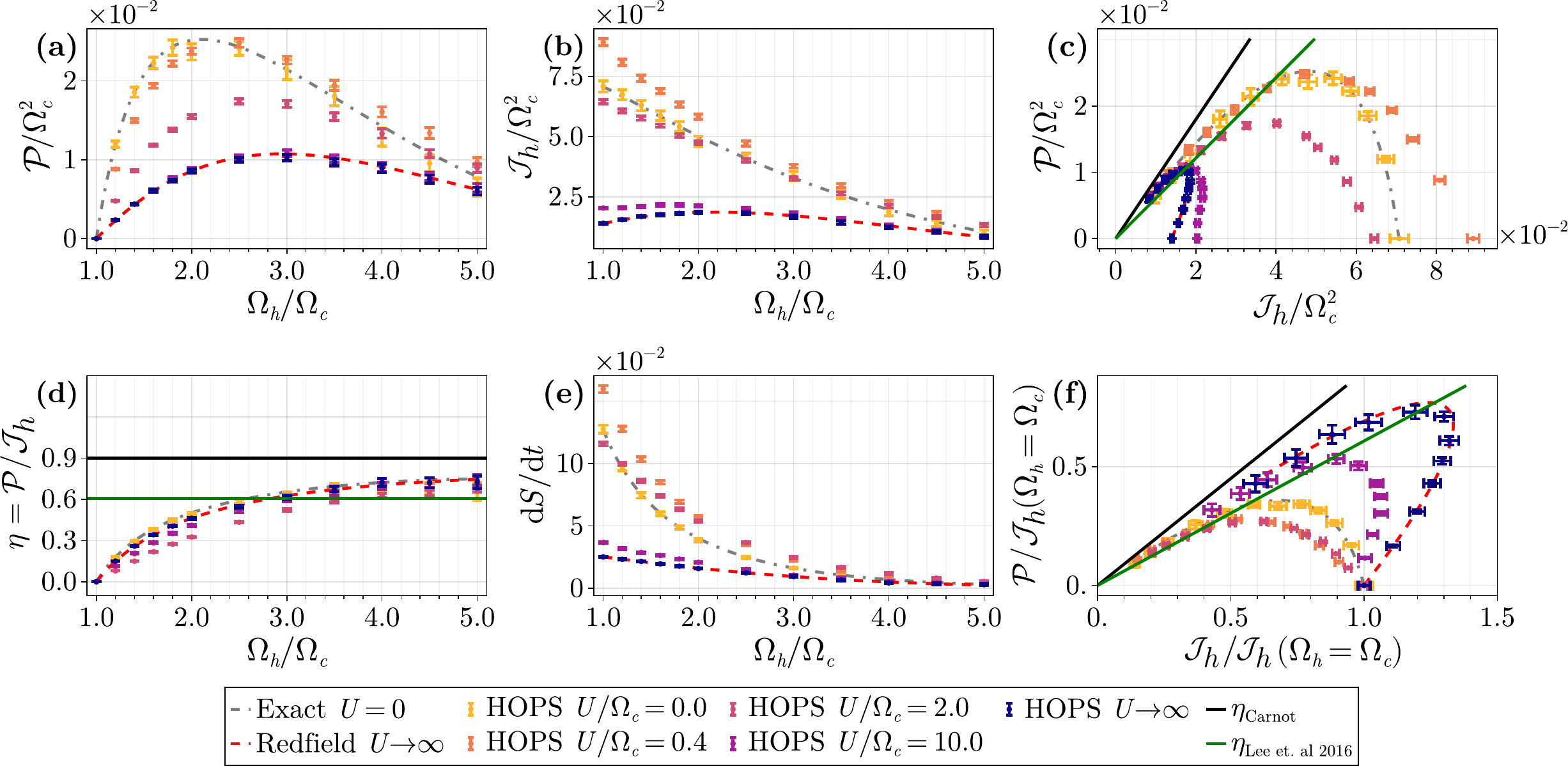}
    \caption{Thermodynamic quantities of the thermal machine at high temperatures.
    (a) Power $\mathcal{P}$ (Eq.~\eqref{def:HOPS:power} for HOPS, \cref{def:power_rf} for Redfield and Exact method) is always positive, and the refrigerator regime vanishes. The value of $\Omega_h/\Omega_c$ where maximum power is produced increases with $U$.
    (b) Heat current from hot bath $\mathcal{J}_h$ (Eq.~\eqref{def:HOPS:currents} for HOPS, Eq.~\eqref{def:currents_exact_rf} for Redfield and Exact method) does not decrease monotonically with increasing $\Omega_h/\Omega_c$ for high interaction strengths $U$. (c) Power $\mathcal{P}$ against heat current $\mathcal{J}_h$ shows that increasing $U$ decreases the magnitudes of thermodynamic quantities. (d) Efficiency $\eta$ (Eq.~\eqref{def:efficiency}) increases with $\Omega_h/\Omega_c$ in the heat engine regime. We observe some difference among the efficiencies for different $U$. HOPS error bars for efficiency are obtained from bootstrapping. (e) Entropy production rate $\mathrm{d}S/\mathrm{d}t$ (Eq.~\eqref{def:entropy}) is always positive, in accordance with the second law of Thermodynamics, and decreases monotonically. (f) Power $\mathcal{P}$ against heat current $\mathcal{J}_h$ rescaled by $\mathcal{J}_h(\Omega_h=\Omega_c)$, but this does not result in the collapse of the curves from (c). The efficiency at maximum power can be read from (f) as the gradient of the straight line connecting the origin to the point on the graph with the highest power. The lines corresponding to the Carnot efficiency and that predicted by \cite{lee_efficiency_2016} are illustrated as straight lines in panels (c), (d), (f).
    Parameters: 
    $T_h/\Omega_c=5.0, T_c/\Omega_c=0.5, \Omega_h/\Omega_c\in[1.0, 5.0]$. $\Omega_c=1.0, \kappa_h=\kappa_c=0.05\Omega_c, g=0.5 \Omega_c$. HOPS parameters are listed in Table~\ref{table:HOPS:params}. Exact method parameters: $N_{\mathrm{bath}}=2000, t_{\mathrm{evo}}=30/\kappa_c$.}
    \label{fig:U:thermo_T_50}
\end{figure*}

Fig.~\ref{fig:U:thermo_T_10} and \ref{fig:U:thermo_T_50} show the effect of $U$ for low and high temperatures, respectively, on the power $\mathcal{P}$, left heat current $\mathcal{J}_h$, efficiency $\eta =\mathcal{P}/\mathcal{J}_h$, and entropy production rate $\mathrm{d}S/\mathrm{d}t$ of the two reservoirs given by
\begin{equation}
    \frac{\mathrm{d}S}{\mathrm{d}t}=-\frac{\mathcal{J}_h}{T_h}-\frac{\mathcal{J}_c}{T_c}.
    \label{def:entropy}
\end{equation}
The junction operates as a heat engine when $\mathcal{P},\mathcal{J}_h>0$ and as a refrigerator when $\mathcal{P},\mathcal{J}_h<0$. 

For low temperatures, increasing $U$ shifts the critical ratio of $\Omega_h/\Omega_c$ separating the heat engine and heat pump/refrigerator regimes, i.e., the stalling point $\lambda_{\mathrm{stall}}$. This can be seen in Fig.~\ref{fig:U:thermo_T_10} (a), (b), where $\mathcal{P}, \mathcal{J}_h$ lines cross zero. This phenomenon can be explained as follows: the non-interacting case ($U\to 0$) under the assumptions of the GME is composed of two independent thermal machines operating at frequencies $\Omega_h\pm g, \Omega_c\pm g$ with efficiencies $\eta_\pm=1-\frac{\Omega_c\pm g}{\Omega_h\pm g}$. As the efficiency of a heat engine approaches the Carnot limit $\eta_C=1-\frac{T_c}{T_h}$, it \textit{stalls}~\cite{hoferMarkovianMasterEquations2017}. The stalling points $\lambda_{\mathrm{stall}\pm}$ thus satisfy $\frac{\Omega_c\pm g}{\Omega_h\pm g}=\frac{T_c}{T_h}$, and the stalling point of the combined thermal machine should lie between $\lambda_{\mathrm{stall}-}$ and $\lambda_{\mathrm{stall}+}$. In the presence of interactions, the system's transition frequencies become $\{\Omega_\ell \pm g + r \frac{U}{2}~|~ r\in \mathbb{N}\}$ in the weak interaction regime $U<g$, and $\{\Omega_\ell \pm g\} \cup\{\Omega_\ell + r U~|~ r\in \mathbb{N}\}$ in the strong interaction regime $U>g$, as described by perturbation theory in App.~\ref{app:perturbation}. Hence, the increase in transition frequencies implies that the overall system will stall at a higher frequency $\Omega_h/\Omega_c$. 

We also observe that for low temperatures, increasing the interaction strength $U$ does not affect the overall shape of the power $\mathcal{P}$ and heat current $\mathcal{J}_h$ -- what we call constitutive relations of the quantum heat engine -- but rather decreases their magnitudes simultaneously for the whole range of system frequencies $\Omega_h/\Omega_c$. Fig.~\ref{fig:U:thermo_T_10} (c) shows that this yields a relatively unaltered efficiency, highlighting the robustness of the thermodynamic property of the junction against on-site interactions. Note that near the stalling point, the efficiencies predicted by HOPS diverge due to division by vanishing quantities. As the magnitudes of the currents are reduced by the on-site interactions, so are the entropy production rates. 

This kind of `universal' behaviour in efficiency at low temperature is further analysed in Fig.~\ref{fig:U:thermo_T_10} (c), displaying the power as a function of the heat current, and in Fig.~\ref{fig:U:thermo_T_10} (f) by rescaling them using the value of the heat current $\mathcal{J}_h(\Omega_h=\Omega_c)$ when the system is a simple quantum junction not producing any power, $\mathcal{P}(\Omega_h=\Omega_c)=0$. 

At high temperatures, however, only the heat engine regime survives. The refrigerator regime would require very high values of $\Omega_h/\Omega_c$, far beyond the bath frequency cut-off $\omega_\mathrm{cut}$. Figure~\ref{fig:U:thermo_T_50} (a), (b), (e) shows that a system with small $U$ can outperform a system with $U=0$ in terms of power, heat current, or entropy production. For large $U$, $\mathcal{P}$ and $\mathcal{J}_L$ surprisingly increase together when $\Omega_h/\Omega_c$ are increased, highlighting a non-trivial non-monotonic behaviour of the thermodynamic quantities as a function of $U$. The collapse of the curves corresponding to different interaction strengths no longer occurs, as seen in Fig.~\ref{fig:U:thermo_T_50} (f). For $U\to\infty$, our results also agree with the expression for universal efficiency at maximum power as given in \cite{lee_efficiency_2016} for the two-qubit case, shown as a straight green line whose gradient is the predicted efficiency. 

\subsection{Non-Reciprocal Transport and Rectification}
\begin{figure}[t]
    \centering
    \includegraphics[width=\linewidth]{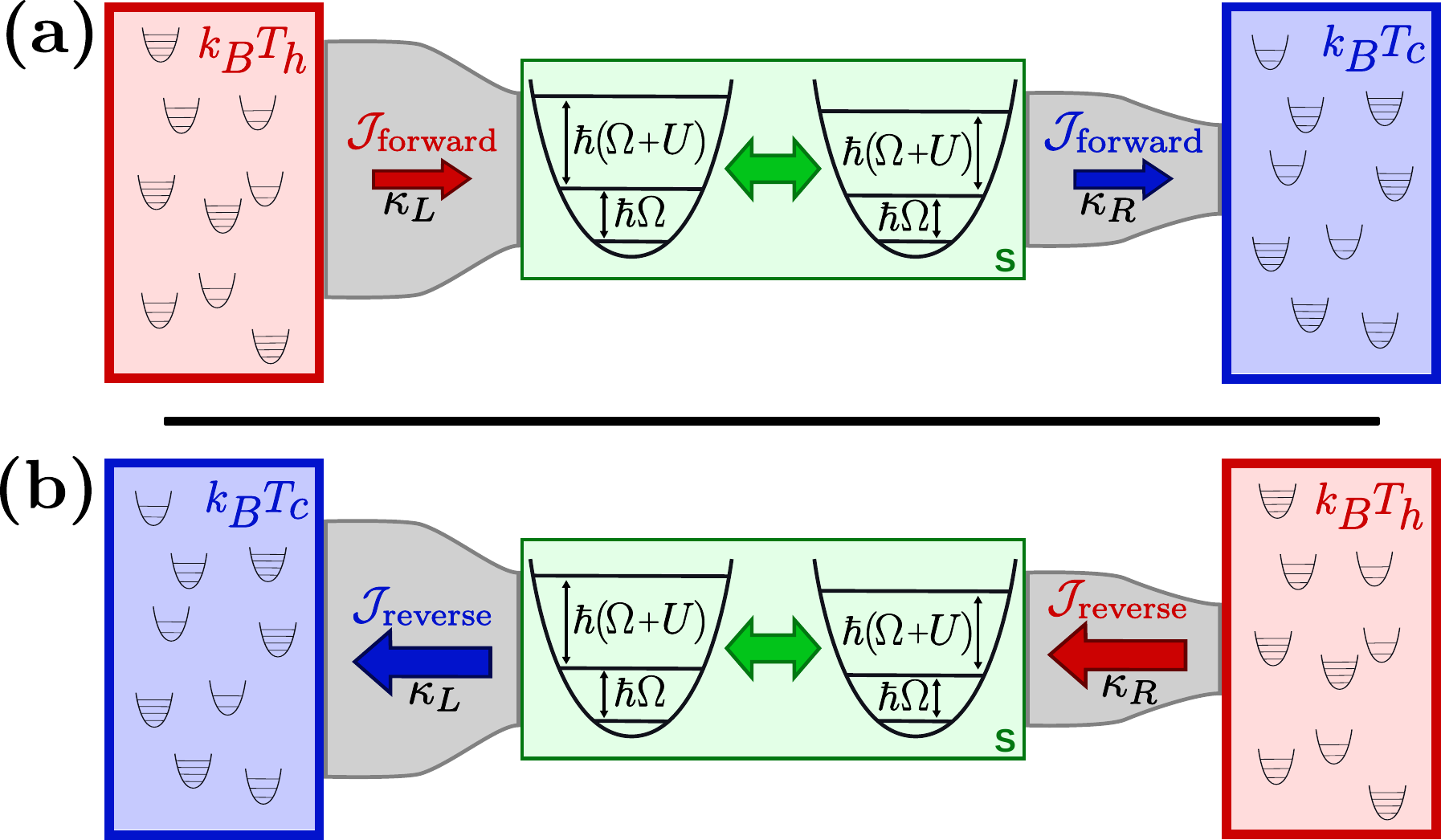}
    \caption{Illustration of the definitions of (a) $\mathcal{J}_{\mathrm{forward}}$, (b) $\mathcal{J}_{\mathrm{reverse}}$, where the width of the grey band represents the magnitude of the junction-reservoir coupling. On top of that, the size of the arrows representing the heat currents illustrates that the reverse setup produces a larger current than the forward one.}
    \label{fig:rect_diagram}
\end{figure}

Beyond the thermodynamic engine properties of the junction studied above, another interesting aspect is the ability of the system to provide non-reciprocal transport when asymmetric tunnelling rates $\kappa_h>\kappa_c$ are introduced, together with the on-site interaction $U$. We focus here on the case of identical system site frequencies $\Omega_h = \Omega_c$, i.e. the system does not experience a time-dependent drive and acts as a passive circuit element. This yields opposite hot and cold heat currents and no power output. In this case, we find that the on-site interaction $U>0$ causes the magnitude of heat currents to depend on the directionality of the temperature bias. To characterize this, we define the rectification ratio
\begin{equation}
    \mathcal{R}:=\left| \frac{\mathcal{J}_{\mathrm{forward}}+\mathcal{J}_{\mathrm{reverse}}}{\mathcal{J}_{\mathrm{forward}}-\mathcal{J}_{\mathrm{reverse}}}\right|,
\label{def:rect}
\end{equation}
\noindent
where $\mathcal{J}_{\mathrm{forward}}$ corresponds to the heat currents $\mathcal{J}_h=-\mathcal{J}_c$ in the standard configuration $\kappa_h>\kappa_c$, $ T_h>T_c$, and where $\mathcal{J}_{\mathrm{reverse}}$ corresponds to the heat currents where the temperatures of the baths are reversed, i.e., $\kappa_h>\kappa_c, T_h<T_c$, as illustrated in Fig.~\ref{fig:rect_diagram}. Note that when computing $\mathcal{R}$ from HOPS trajectory data, in order to minimise its variance, we choose to define the heat currents $\mathcal{J}_{\mathrm{forward}}$ and $\mathcal{J}_{\mathrm{reverse}}$ as the heat current from the reservoir that has a smaller $\kappa_\ell T_\ell$, which acts as an estimate of the magnitude of trajectory fluctuations. 

For a perfect diode, $\mathcal{R}=1$ since one of the directions does not allow current to pass through. Any $\mathcal{R}>0$ gives potential for the junction to be used as a logical element.

\begin{figure}[t]
    \centering
\includegraphics[width=\linewidth]{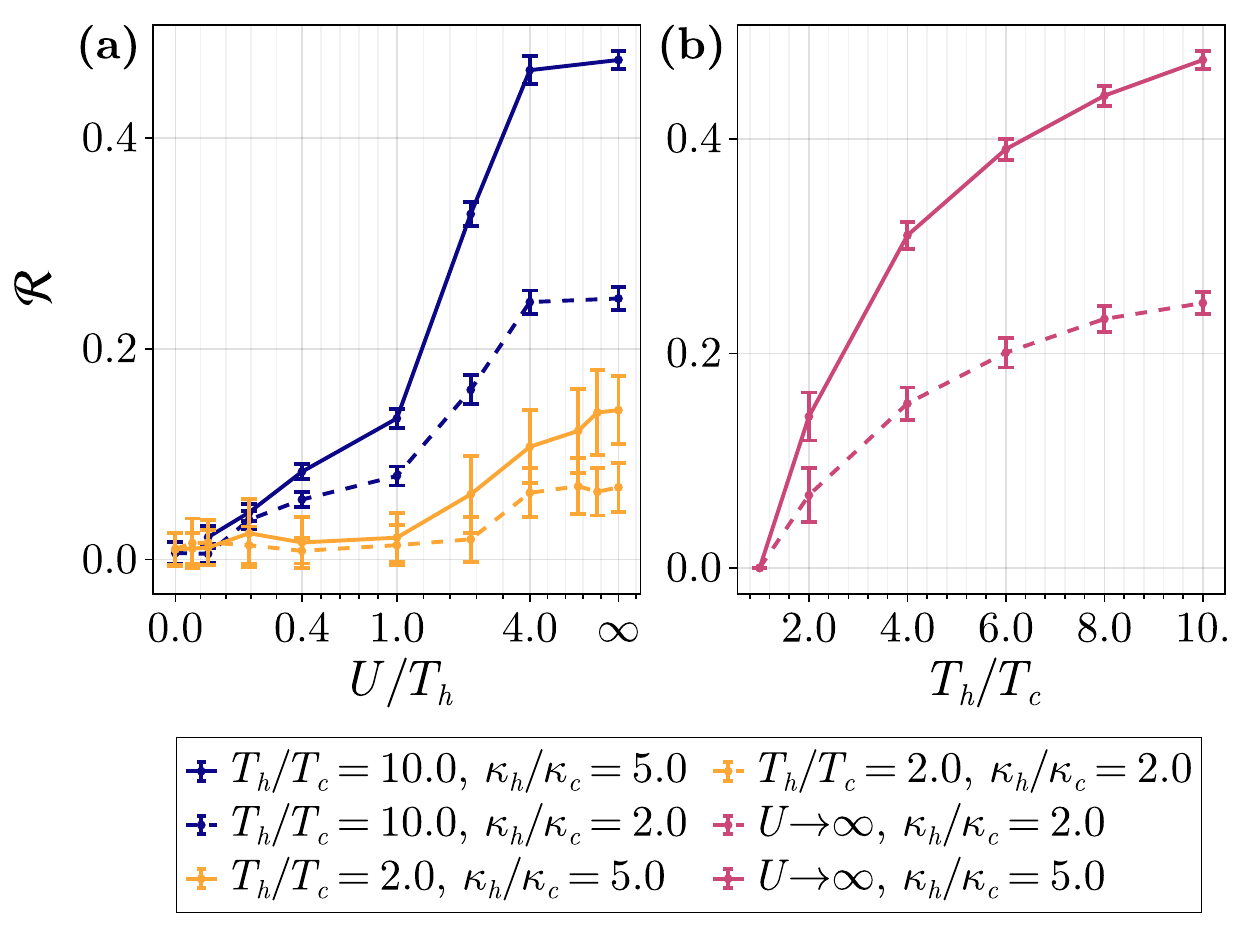}
\caption{Rectification ratio $\mathcal{R}$ (Eq.~\eqref{def:rect}) calculated using HOPS. (a) Rectification increases with $U$, and reaches an asymptotic value at approximately $U/T_h=4.0$. (b) Rectification for two qubits ($U\to\infty$) increases with both $T_h/T_c$ and $\kappa_h/\kappa_c$. System parameters: $\Omega_h/\Omega_c = 1, g/\Omega_c = 0.5, \kappa_c/\Omega_c = 0.05, T_c/\Omega_c=0.5$. HOPS parameters are listed in Table~\ref{table:HOPS:params}.} 
    \label{fig:U:HOPS:non_reciprocal_U}
\end{figure}

In the harmonic oscillator case ($U=0$), we find that rectification is negligible, $\mathcal{R}\sim10^{-4}$. It becomes exactly zero when the Lamb shifts are neglected or in the weak coupling limit $\kappa_{h,c}\to 0$. In contrast, for $U > 0$, $\mathcal{R}$ increases with $U$, $T_h/T_c$, and $\kappa_h/\kappa_c$, as can be seen in Fig.~\ref{fig:U:HOPS:non_reciprocal_U}. In particular, for the parameters considered, we observe that $|\mathcal{J}_{\mathrm{forward}}|<|\mathcal{J}_{\mathrm{reverse}}|$, which means that the system conducts more heat current if the stronger coupling (i.e., the smaller potential barrier) is on the cold side. Note that in the limit $U\to \infty$, when the junction consists of two qubits, similar conclusions can be extended to a chain of $N$ qubits~\cite{santiago-garcia_quantum_2025}. The asymmetric coupling of the baths is a key requirement for rectification \cite{QuantumThermalTrans}; consequently, setting $\kappa_h = \kappa_c$ automatically leads to zero rectification for all values of $U$. The same result is found when considering the thermal equilibrium case, $T_h = T_c$. 
\subsection{Entanglement Generation}
\begin{figure}[t]
    \centering
\includegraphics[width=0.97\linewidth]{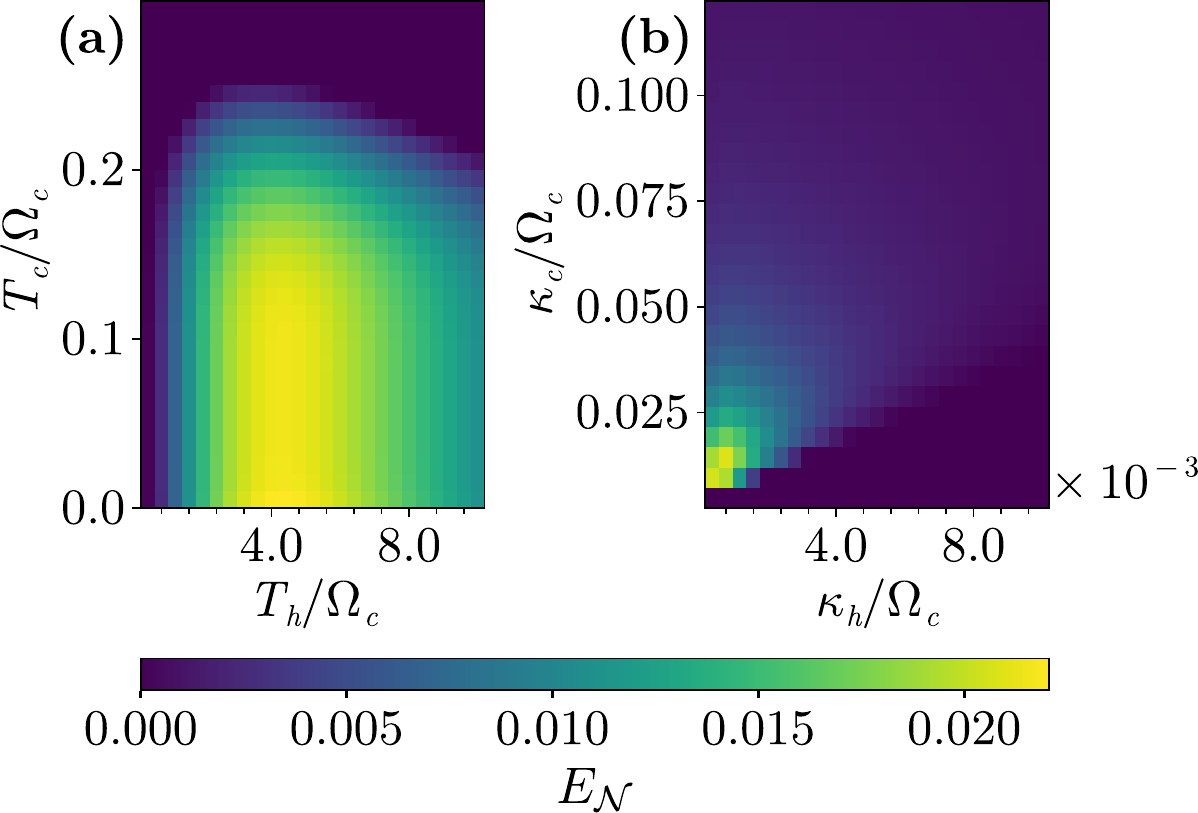}
    \caption{Logarithmic negativity $E_\mathcal{N}$ as a function of (a) bath temperatures and (b) coupling to the baths. In (a), $E_\mathcal{N}$ remains positive only below $T_c/\Omega_c\approx 0.24$ and for $T_h\gg T_c$. In (b), positivity is strictly restricted to the regime where $\kappa_c\gg\kappa_h,g$. 
    System parameters: $\Omega_h/\Omega_c=1.0, g/\Omega_c=0.0025$, and (a) $\kappa_h/\Omega_c=0.008, \kappa_c/\Omega_c=0.015$; (b) $T_h/\Omega_c=5.4, T_c/\Omega_c=0.05$.}
    \label{fig:log_neg}
\end{figure}
We now discuss how entanglement in the junction can be generated and controlled by the parameters of the system. We find that asymmetry in the bath coupling $\kappa_h\ne\kappa_c$ is necessary to create steady-state entanglement between the two sites, in addition to suitably small coupling strengths $\kappa_h, \kappa_c, g \ll \Omega_{h},\Omega_c$. Log-negativity $E_\mathcal{N}$ is a measure of entanglement \cite{vidal_computable_2002} compatible with concurrence~\cite{verstraete_comparison_2001}, which was previously investigated in the case of two-qubits~\cite{bohr_brask_autonomous_2015}. 
Fig.~\ref{fig:log_neg} shows log-negativity  $E_\mathcal{N}$ as a function of the  temperatures $T_h$ and $T_c$ and of the tunnelling rates $\kappa_h$ and $\kappa_c$ in the case of two identical qubits ($U \to \infty$) and $\Omega_h = \Omega_c$. In contrast, we did not find any entanglement for $U = 0$ within the investigated parameter regimes, suggesting that on-site interactions constitute a powerful knob to generate entanglement in transport junctions. Note that in \cite{goold_role_2016}, it was suggested that starting from a single thermal state, there is a maximum temperature below which entanglement can be generated. We observe a maximum of $T_c/\Omega_c\approx 0.24$, in comparison with the results $T_c/\Omega_c\approx 0.21$ obtained by \cite{bohr_brask_autonomous_2015} under Markovian approximations.

From a method perspective, we note that in this parameter regime, the small coupling strengths make HOPS an inefficient method to obtain the steady-state density matrix as the required convergence time $t_{\mathrm{evo}}$ and number of time-steps are very large. On the other hand, RME is accurate and efficient in this regime and was thus used to generate Fig.~\ref{fig:log_neg}.

\begin{figure}[t]
    \centering
\includegraphics[width=\linewidth]{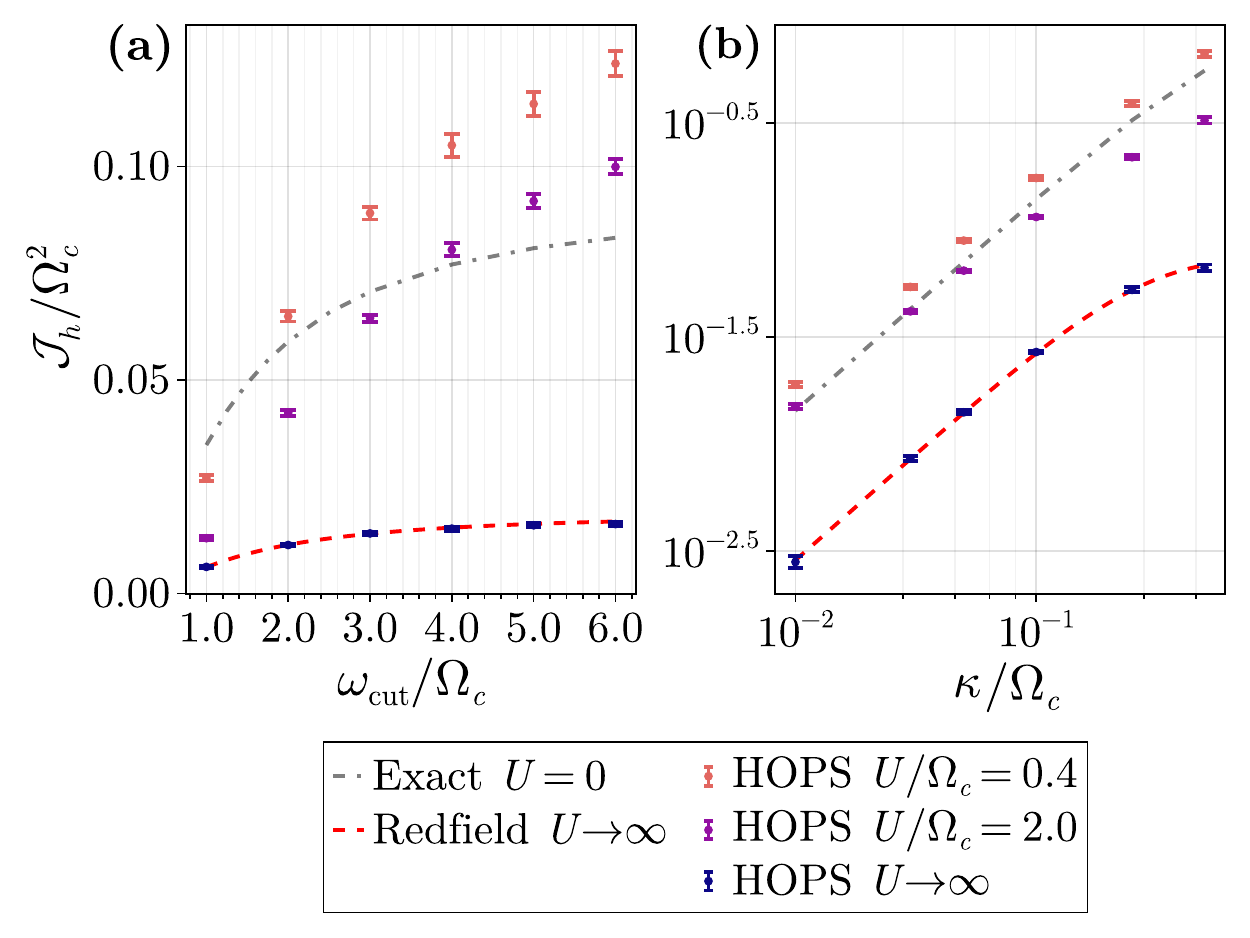}

    \caption{(a) Heat current $\mathcal{J}_h$ (Eq.~\eqref{eq:def_currents_HOPS}) increases with bath cut-off frequency $\omega_\mathrm{cut}/\Omega_c\in[1.0,6.0]$. The asymptotic value of $\mathcal{J}_h$ is higher for $U>0$ than both $U=0$ and $U\to \infty$. (b) Heat current $\mathcal{J}_h$ follows a power law relationship with coupling strength $\kappa_c$. 
    System parameters: $T_h/\Omega_c=5.0, T_c/\Omega_c=0.5, \kappa_{h}=\kappa_c=\kappa, \Omega_h/\Omega_c=1.0, g/\Omega_c=0.5$. (a) $\kappa/\Omega_c=0.05$. (b) $\omega_\mathrm{cut}/\Omega_c=3.0$. HOPS parameters are listed in Table~\ref{table:HOPS:params}. Exact method parameters: $N_{\mathrm{bath}}=2000, t_{\mathrm{evo}}=30/\kappa$.} 
    \label{fig:U:omega_c}
\end{figure}
\begin{figure*}[!ht]
    \centering
\includegraphics[width=0.95\linewidth]{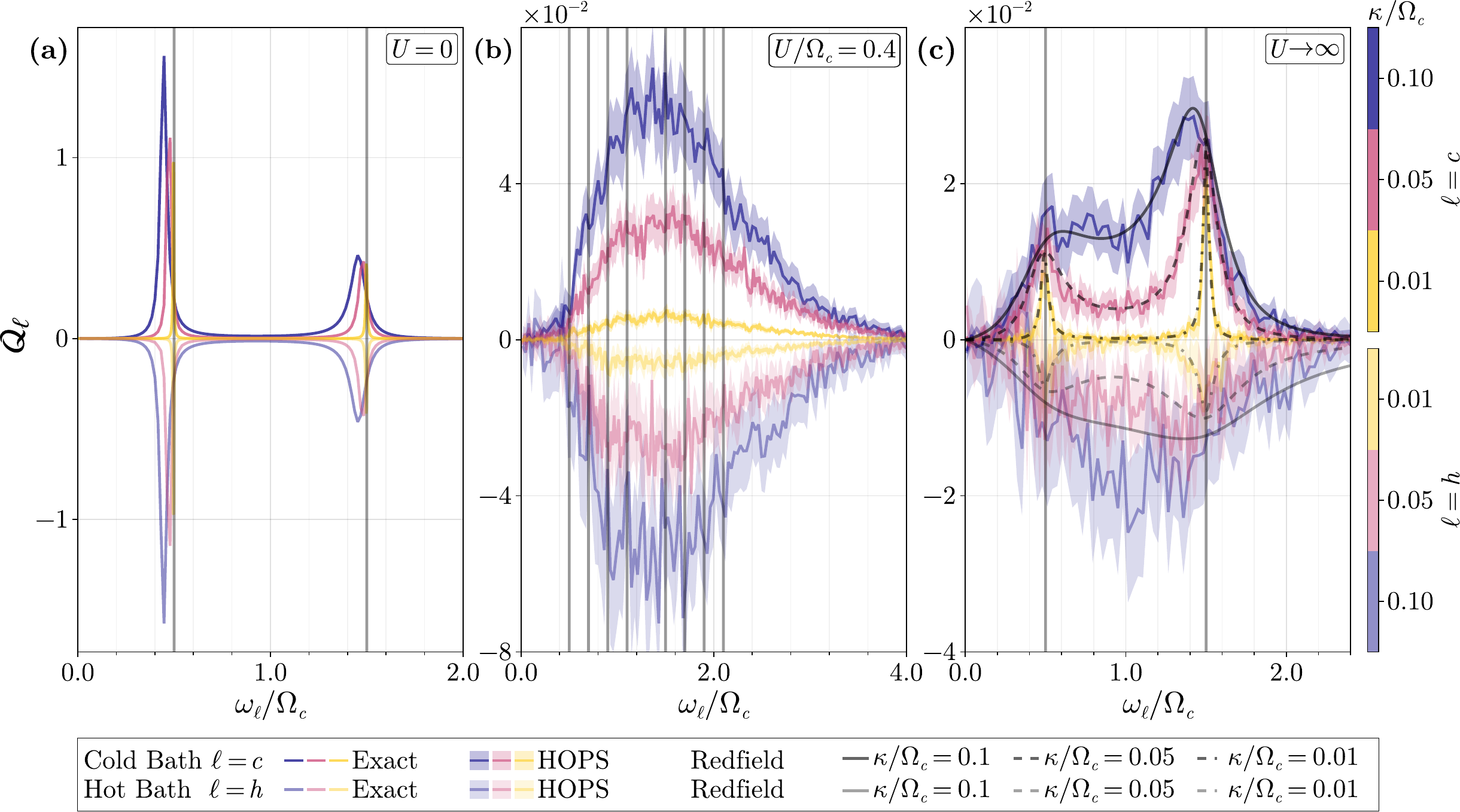}
    \caption{Momentum-resolved particle currents $\mathcal{Q}_\ell$ (Eq.~\eqref{def:particle_current}). Positive values represent particle flow into the bath $\ell$, and vice versa. Vertical grey lines indicate up to the first four transition frequencies as predicted by perturbation theory (App.~\ref{app:perturbation}), showing good agreement with peaks at weak system-bath coupling strengths. (a) For $U=0$, there are two Lorentzian peaks for the system eigenfrequencies $\Omega_\ell \pm g$. They broaden while retaining their shapes as coupling strength $\kappa$ is increased. They are displaced to the left due to Lamb shift. (b) For $U>0$, at very small $\kappa$ one can observe a series of peaks corresponding to eigenfrequencies as described in App.~\ref{app:perturbation}. As $\kappa$ increases, the peaks merge into a larger shape. They increase in magnitude with $\kappa$. (c) For $U\to\infty$, the system returns to only having eigenfrequencies $\Omega_\ell \pm g$. The peaks broaden at a much faster rate than for the $U=0$ case in (a), and the overall magnitude does not grow as is the case in (b). $\mathcal{Q}_h$ and $\mathcal{Q}_c$ are notably asymmetrical at large $\kappa$.
    System parameters: $T_h/\Omega_c=5, T_c/\Omega_c=0.5, \kappa_h=\kappa_c=\kappa, \Omega_h/\Omega_c=1.0,g/\Omega_c=0.5$. HOPS parameters are listed in Table~\ref{table:HOPS:params}. Exact method parameters: $N_{\mathrm{bath}}=1000, t_{\mathrm{evo}}=20/\kappa_{h,c}$. 
    }
    \label{fig:U:HOPS:broadening}
\end{figure*}

\subsection{Reservoir Engineering}
We now turn to the effects of reservoir engineering on the transport properties of the junction. The bath spectral density (Eq.~\eqref{eq:ohmic}) depends on two parameters for its shape, the cut-off frequency $\omega_{\mathrm{cut}}$ and the coupling strength $\kappa_\ell$. In Fig.~\ref{fig:U:omega_c}, we investigate the effect of varying each parameter on the heat current $\mathcal{J}_h$ in the case $\Omega_h=\Omega_c$, where the system does not output power.

When $U$ is small, there is a spread of system eigenfrequencies forming a ladder starting from $\Omega_\ell\pm g$ as predicted by perturbation theory (App.~\ref{app:perturbation}). They experience increased coupling strength due to the Ohmic spectral density (Eq.~\eqref{eq:ohmic}), up to the cut-off frequency $\omega_\mathrm{cut}$. As shown in Fig. \ref{fig:U:omega_c} (a), the heat current $\mathcal{J}_h$ increases more rapidly for $U>0$ than for the harmonic oscillator ($U=0$) and the qubit ($U\to\infty$) case. 
Increasing the coupling strength $\kappa$ leads to a power law growth of heat current $\mathcal{J}_h$ for $U=0$. For $U>0$, deviations from power law appears at strong coupling $\kappa_c$. For the qubit case, the magnitude of $\mathcal{J}_h$ is damped by almost an order of magnitude. The presence of on-site interactions $U$ suppresses both the number of Fock states in the strong coupling regime and the evolution time $t_{\mathrm{evo}}$ required in the weak coupling regime, thereby enabling the use of HOPS in Fig.~\ref{fig:U:omega_c} for $U\ne 0$.

\subsection{Momentum-Resolved Particle Currents}\label{sec:spectrum}

Fig.~\ref{fig:U:HOPS:broadening} shows the effect of increasing interaction strength $U$ and coupling strength $\kappa$ on the momentum-resolved particle current spectrum. One can identify peaks centered around the system's eigenfrequencies (indicated by vertical grey lines) for $U=0,\infty$ and for $U>0$ when $\kappa/\Omega_c\ll 1$ is small. As can be seen in the figure, the effect of $U$ consists in first splitting the initial two peaks into multiple ones that quickly merge into a wide crest, before resolving, for large $U$, to two broad peaks. The non-Markovian nature of the baths is evident in the broad and unequal spectrum of frequencies participating in transport, and that the dominant frequencies can be modified by different parameters. The impact of non-linearity of the system's interaction is also most evident in the two-qubit case (Fig.~\ref{fig:U:HOPS:broadening} (c)), where the transport spectra for the hot and cold baths are not symmetrical.

\section{Conclusion and Outlook}
\label{sec:conclusion}

We have studied an interacting two-site bosonic thermodynamic junction and shown that the Redfield master equation accurately captures its steady-state observables and transport properties in regimes where the local and global approaches break down. Benchmarking against exact dynamics in the non-interacting limit demonstrates that the Redfield description interpolates between these two limiting reduced approaches while remaining quantitatively reliable over a broad range of parameters. 

Going beyond the weak-coupling regime with HOPS, we have shown that on-site interactions provide a versatile control knob for thermodynamic functionality. At low temperatures, interactions largely rescale the heat current and power while preserving their constitutive relation, leading to an emergent universal-like robustness of the efficiency. At higher temperatures, this robustness is lost, and interactions can instead be leveraged to tune the thermodynamic performance of the machine. Under asymmetric reservoir couplings, the same non-linearities enable non-reciprocal transport, and in suitable parameter regimes they also generate steady-state entanglement within the junction. These results show that interaction-induced nonlinearity offers a route to functional thermodynamic behaviour distinct from architectures relying on permanent structural asymmetry~\cite{magazzu_thermal_2025}.

Beyond macroscopic thermodynamic observables such as heat currents, power, and efficiency, we have also identified microscopic signatures of the underlying transport mechanisms. Momentum-resolved particle-current spectra, which are in principle measurable quantities, display clear imprints of both interaction-induced nonlinearities and non-Markovian reservoir effects. They therefore provide an experimentally relevant route to diagnose the mechanisms underpinning the transport phenomena reported here and to support their interpretation at the level of reservoir-resolved currents.

Finally, the Redfield master equation and HOPS offer complementary descriptions. The former relies on weak system-reservoir couplings, whereas the latter does not, so that each method is advantageous in different parameter regimes in terms of accuracy and computational cost. A natural perspective is to extend the regime of validity of reduced master-equation approaches by incorporating corrections arising from stronger coupling to the baths~\cite{PhysRevB.107.195117,PhysRevLett.120.120602}, thereby increasing our capability to model dissipation and non-equilibrium dynamics beyond weak coupling~\cite{correa_enhancement_2017}. 

A further natural direction is to include dissipative channels acting directly within the junction, in addition to the thermal reservoirs considered here. Dissipation is well known to provide a means of controlling transport: dephasing-assisted transport has been widely studied in non-interacting or effectively single-excitation settings~\cite{Plenio2008,Rebentrost2009,Caruso2009,Chin2010,Lacerda2021}, while local particle loss and dephasing can also be exploited in genuinely interacting junctions~\cite{Damanet2019}. In lattice systems, local losses can give rise to symmetry-protected transport~\cite{Visuri2022} and strongly reshape nonlinear current-voltage characteristics~\cite{Visuri2023}; more general transport formulae have also been developed for interacting systems subject to Markovian gain and loss processes~\cite{Jin2020}. A closely related perspective is continuous monitoring of the junction: measurement backaction can induce non-reciprocal transport and enable work extraction~\cite{Ferreira2024}, while cavity-based proposals show that atomic currents may be monitored continuously with controlled backaction~\cite{Uchino2018}. These directions are particularly natural in cold-atom settings~\cite{Corman2019,Huang2023}, but are also increasingly relevant to trapped ions, where spin-boson transfer models with independently tunable coherent and dissipative processes have recently been demonstrated~\cite{So2024}, and to superconducting circuits, where tunable dissipation of bosonic resonators and photonic heat transport through non-linear circuit elements are already experimentally accessible~\cite{Sevriuk2019,Gubaydullin2022}.

\begin{acknowledgments}
The authors would like to acknowledge the use of the University of Oxford Advanced Research Computing (ARC) facility in carrying out this work (https://doi.org/10.5281/zenodo.22558)
Computational resources have also been provided by the Consortium des Équipements de Calcul Intensif (CÉCI), funded by the Fonds de la Recherche Scientifique de Belgique (F.R.S.-FNRS) under Grant No. 2.5020.11 and by the Walloon Region.

\end{acknowledgments}

\appendix

\section{Master Equations}\label{app:ME}

\subsection{Steady State Solutions for $U = 0$}

For non-interacting particles ($U = 0$) and weak junction-reservoir coupling, all master equations derived in the main text are quadratic and can be solved \textit{analytically} using e.g. third quantization~\cite{prosenThirdQuantizationGeneral2008, prosenQuantizationBosonOperator2010} or covariant matrix/cumulant expansion methods. 

Using third quantization, for the LME (Eq.~\eqref{driving local}), we obtain the following steady-state correlations
\begin{align}\label{correlationLocal}
    \langle a_h^\dagger a_h \rangle^L_\mathrm{SS} &=n^h - \frac{g^2(J_h+J_c)J_c[n_h-n_c]}{(J_h+J_c)^2(\pi^2 J_h J_c +g^2) +J_hJ_c \Sigma^2}\nonumber,\\
     \langle a_c^\dagger a_c \rangle^L_\mathrm{SS} &=n^c - \frac{g^2(J_h+J_c)J_h[n_c-n_h]}{(J_h+J_c)^2(\pi^2 J_h J_c +g^2) +J_hJ_c\Sigma^2}\nonumber,\\
      \langle a_h^\dagger a_c \rangle^L_\mathrm{SS} &= \frac{gJ_hJ_c(n_h-n_c)[\mathrm{i}\pi(J_h+J_c)+\Sigma]}{(J_h+J_c)^2(\pi^2 J_h J_c +g^2) +J_hJ_c\Sigma^2},
\end{align}
where $\Sigma = \sum_\ell\xi_\ell(\bar{\Delta}_\ell+\Delta_\ell)$ is a combination of the different Lamb shifts that appears in the master equation. Neglecting this correction, we retrieve the results provided in \cite{hoferMarkovianMasterEquations2017}. We understand the expression for the hot (cold) population as the one obtained for a thermal Gibbs state (which would be the corresponding steady state for a single site coupled to only one bath) modified by a correction due to the interaction with the other site.

Considering the correlations at the steady state associated with the GME (Eq.~\eqref{driving global}), it is more convenient to first write them in the system Hamiltonian eigenbasis (i.e., the $\{a_\pm,a^\dagger_\pm\}$). We obtain
\begin{equation}\label{correlationGlobal}
   \langle a^\dagger_\pm a_\pm \rangle^G_\mathrm{SS} = \frac{J_{h\pm } n^{h \pm} +J_{c \pm} n^{c \pm}}{J_{h\pm} +J_{c \pm}}.
\end{equation}
For the cross-terms, $\langle a^\dagger_\pm a_\mp\rangle^G = 0$ due to our choice of basis, as the $+$ and $-$ modes are uncoupled. Therefore, it is trivial to calculate the correlations in the $\{a_\ell,a^\dagger_\ell\}$ basis, we find 
\begin{equation}
    \langle a^\dagger_\ell a_{\bar{\ell}}\rangle^G_\mathrm{SS} = \frac{\langle a_-^\dagger a_- \rangle^G_\mathrm{SS}+\xi_\ell \xi_{\bar{\ell}} \langle a_+^\dagger a_+ \rangle^G_\mathrm{SS}}{2},
\end{equation}
meaning that we find the same population for each site and that the cross-term has no imaginary part. Notably, we see that these expressions do not contains the Lamb shifts corrections. It is also worth noting that the results obtained by the global master equation are independent of the parameter $\kappa_\ell$ when taken identical for hot and cold reservoirs.

Finally, we now present the results for the RME (Eq.~\eqref{Redfieldstd2}). Since the full analytical expressions including Lamb shifts are too cumbersome to display here, we present the simplified case where they are neglected. Using this approach, the correlations can be written as a sum of the corresponding GME correlation and a correction term, which accounts for the effective coupling between the $+$ and $-$ modes induced by the counter-rotating dissipative terms in the RME. We find
\begin{widetext}
\begin{align}
    \langle a_+^\dagger a_+ \rangle^R_\mathrm{SS} = &\langle a_+^\dagger a_+ \rangle^G_\mathrm{SS}\nonumber \\
    &- \frac{(J_{h-} - J_{c-}) (J_{h-} + J_{h+} + J_{c-} + J_{c+})\pi^2[J_{h+} J_{c+} (J_{h-} + J_{c-}) (n^{h+} - n^{c+})+J_{h-} J_{c-} (J_{h+} + J_{c+}) (n^{h-} - n^{c-})]}{(J_{h+}+J_{c+})[8 g^2 (J_{h-} + J_{c-}) (J_{h+} + J_{c+})+(J_{h-} + J_{h+} + J_{c-} + J_{c+})^2 (J_{h+} J_{c-} + J_{h-} J_{c+}) \pi^2]}, \nonumber\\
    \langle a_+^\dagger a_- \rangle^R_\mathrm{SS} =&  \frac{\pi[J_{h+} J_{c+} (J_{h-} + J_{c-}) (n^{h+} - n^{c+})+J_{h-} J_{c-} (J_{h+} + J_{c+}) (n^{h-} - n^{c-})](4 \mathrm{i} g - \pi[J_{h-} + J_{h+}  + J_{c-}  + J_{c+} )]}{8 g^2 (J_{h-} + J_{c-}) (J_{h+} + J_{c+})+(J_{h-} + J_{h+} + J_{c-} + J_{c+})^2 (J_{h+} J_{c-} + J_{h-} J_{c+}) \pi^2},\nonumber\\
    \langle a_-^\dagger a_- \rangle^R_\mathrm{SS} =& \langle a_-^\dagger a_- \rangle^G_\mathrm{SS}\nonumber\\
    &- \frac{(J_{h+} - J_{c+}) (J_{h-} + J_{h+} + J_{c-} + J_{c+})\pi^2[J_{h+} J_{c+} (J_{h-} + J_{c-}) (n^{h+} - n^{c+})+J_{h-} J_{c-} (J_{h+} + J_{c+}) (n^{h-} - n^{c-})]}{(J_{h-}+J_{c-})[8 g^2 (J_{h-} + J_{c-}) (J_{h+} + J_{c+})+(J_{h-} + J_{h+} + J_{c-} + J_{c+})^2 (J_{h+} J_{c-} + J_{h-} J_{c+}) \pi^2]}.
\end{align}
\end{widetext}

\section{Exact Method -- Heisenberg Equations of Motion}\label{Appendix_exactmethod}

\subsection{Exact Method Formalism}\label{app:exact:formalism}
We first discretise our baths according to the method introduced in \cite{rivas_markovian_2010}, by fixing a number $N_{\mathrm{bath}}$ of harmonic oscillators in each bath such that the frequencies $\omega_{\ell k}$ in the baths' Hamiltonians (Eq.~\eqref{Ham_Baths}) now read
\begin{equation}
\omega_{\ell k} = k\Delta\omega,
\end{equation}
for $k~\in \{1,..., N_{\mathrm{bath}}\}$ with frequency gap $\Delta\omega = 2 \pi/t_\mathrm{evo}$, $t_\mathrm{evo}$ being the total time of evolution considered. This ensures that recurrence does not happen during $t\in[0,t_\mathrm{evo}]$. By making $t_\mathrm{evo}$ sufficiently large, the system approaches its steady state at large times. 

Regarding the interaction Hamiltonian, the coupling coefficients $g_{\ell k}$ are determined through the bath spectral density (Eq.~\eqref{def:bath_spectral_density}). Integrating over frequency space, we obtain
\begin{equation}
    \int_0^\infty \mathrm{d}\omega~ J_\ell (\omega) = \sum_{k=0}^{N_\mathrm{bath}}g_{\ell k}^2.
\end{equation}
By truncating the integral and discretizing it, we obtain the coefficients
\begin{equation}\label{def:glk1}
    g_{\ell k} = \sqrt{J_\ell(\omega_{\ell k})\Delta \omega},
\end{equation}
which are chosen to be real. The Heisenberg equations of motion for all the annihilation operators read
\begin{align}
    i \dot a_h &= g a_c + \sum_k g_{h k}a_{h k}, \nonumber\\
     i \dot a_c &= g a_h + \sum_k g_{c k}a_{c k}, \nonumber\\
     i \dot a_{\ell k} &= (\omega_{\ell k}-\Omega_\ell) a_{\ell k} + g_{\ell k} a_\ell,  \nonumber
\end{align}
This system of coupled equation can be rewritten as
\begin{equation}
    \i \dot{A} = W A,
\end{equation}
where $A= (a_h, a_{h1},\dotsc , a_{hN_{\mathrm{bath}}}, a_{c}, a_{c1},\dotsc , a_{cN_{\mathrm{bath}}})^T$ and where $W$ is the matrix
\begin{widetext}
\begin{equation}
W =
    \begin{pmatrix}
         0 & g_{h1} & \dotsc & g_{hN_{\mathrm{bath}}} & g  \\
         g_{h1} & \omega_{h1}-\Omega_h\\
         \vdots & &\ddots\\
         g_{hN_{\mathrm{bath}}}  & & & \omega_{hN_{\mathrm{bath}}}-\Omega_h\\
         g & & & & 0 & g_{c1} & \dotsc & g_{cN_{\mathrm{bath}}}\\
           & & & & g_{c1} & \omega_{c1}-\Omega_c \\
           & & & & \vdots & & \ddots \\
           & & & & g_{cN_{\mathrm{bath}}} & & & \omega_{cN_{\mathrm{bath}}}-\Omega_c
     \end{pmatrix}.
\end{equation}
\end{widetext}
The solution of this system is given by 
\begin{equation}
    A(t) = T(t) A(0), \quad T(t) = e^{-\i W t}.
\end{equation}
The quantities that will be compared, namely the steady state power and currents, require the computation of the correlation matrix elements at large times. To this end, we find their values at time $t_f=0.95 t_\mathrm{evo}$, given by
\begin{equation}\label{def:corr_elements_exact}
    \langle A_i^\dagger A_j \rangle_{SS} :=\sum_{l,m} T^\dagger_{i l}(t_f)T_{jm}(t_f) \langle A^\dagger_l A_m \rangle(0),
\end{equation}
which is independent of the choice of the initial state, since in the setup that we are interested in, the steady state is unique. The factor of $0.95$ is added to avoid the onset of recurrence at $t=t_{\mathrm{evo}}$.

Here, we chose an initial vacuum state for the junction and thermal states at temperature $T_\ell$ for each bath $\ell$, which has no first order moments $\braket{A(0)}=\mathbf{0}$, which implies $\braket{A(t)}=\mathbf{0}$ as well. 

\subsection{Fidelity Calculation}\label{app:exact:fidelity}
A Gaussian two mode system is defined by its first and second order moments. We use the following formalism to calculate fidelity \cite{hoferMarkovianMasterEquations2017, marian_uhlmann_2012}. 

For the two baths $\ell = h,c$, let $x_\ell = \frac{1}{\sqrt{2 \Omega_\ell}}(a_\ell^\dagger + a_\ell)$, $p_\ell = i\sqrt{\frac{\Omega_\ell}{2}}(a_\ell^\dagger - a_\ell)$. That is to say, the change of basis for the two vectors of operators is
\begin{equation}
    \mathbf{A} = \begin{pmatrix} a_h \\ a_h^\dagger \\ a_c \\ a_c^\dagger \end{pmatrix}, \quad
    \mathbf{B} = \begin{pmatrix} x_h \\ p_h \\ x_c \\ p_c \end{pmatrix}, \quad
    \mathbf{B} = \Lambda \mathbf{A},
\end{equation}
\noindent
where
\begin{equation}
    \Lambda = \begin{pmatrix}
        \frac{1}{\sqrt{2 \Omega_h}} & \frac{1}{\sqrt{2 \Omega_h}} & 0 & 0 \\
        -i\sqrt{\frac{\Omega_h}{2}} & i\sqrt{\frac{\Omega_h}{2}} & 0 & 0 \\
        0 & 0 & \frac{1}{\sqrt{2 \Omega_c}} & \frac{1}{\sqrt{2 \Omega_c}} \\
        0 & 0 & -i\sqrt{\frac{\Omega_c}{2}} & i\sqrt{\frac{\Omega_c}{2}} \\  
    \end{pmatrix}.
\end{equation}

Center the observables as follows, $a_\ell \to a_\ell - \braket{a_\ell}$, $x_\ell \to x_\ell - \braket{x_\ell}$, and $p_\ell \to p_\ell - \braket{p_\ell}$. Then $ \mathbf{B} = \Lambda ~ \mathbf{A}$ still holds. Define the following matrix using creation and annihilation operators,
\begin{equation}
    \Theta = \begin{pmatrix}
        \braket{ a_h^\dagger  a_h} + \frac{1}{2} & \braket{ a_h  a_h} & \braket{ a_h  a_c^\dagger} & \braket{ a_h  a_c} \\
        \braket{ a_h^\dagger  a_h^\dagger} & \braket{ a_h^\dagger  a_h} + \frac{1}{2}& \braket{ a_h^\dagger  a_c^\dagger} & \braket{ a_h^\dagger  a_c} \\
        \braket{ a_h^\dagger  a_c} & \braket{ a_h  a_c} & \braket{ a_c^\dagger  a_c} + \frac{1}{2} & \braket{ a_c  a_c} \\
        \braket{ a_h^\dagger  a_c^\dagger} & \braket{ a_h  a_c^\dagger} & \braket{ a_c^\dagger  a_c^\dagger} & \braket{ a_c^\dagger  a_c} + \frac{1}{2}
    \end{pmatrix}.
\end{equation}

We then obtain the covariance matrix by transforming to the $x,p$ basis,
\begin{equation}
    C = \Lambda \Theta \Lambda^\dagger.
\end{equation}

For two Gaussian two mode systems, let $\Delta_\mathbf{B}=\braket{\mathbf{B}_1}-\braket{\mathbf{B}_2}$ be the difference in their first order moments. Their fidelity is given by \cite{marian_uhlmann_2012}
\begin{equation}
\begin{split}
    \mathcal{F}&(C_1, C_2)  =\exp\left[-\frac{1}{2}\Delta_\mathbf{B}^T(C_1+C_2)^{-1}\Delta_\mathbf{B}\right] \\ & \times\left[ \sqrt{\Lambda_b} + \sqrt{\Lambda_c} - \sqrt{(\sqrt{\Lambda_b} + \sqrt{\Lambda_c})^2 - \Lambda_a}\right]^{-1},
\end{split}
\end{equation}
\noindent
where $\Lambda_a = \det(C_1 + C_2)$, $\Lambda_b = 2^4 \det(J C_1 J C_2 - \mathds{1}/4)$, and
$\Lambda_c = 2^4 \det(C_1 + i J/2) \det(C_2 + i J/2)$. The symplectic matrix is $J=\bigoplus_{\ell=h,c} \begin{pmatrix}
    0 & 1 \\
    -1 & 0
\end{pmatrix}$ .

In this work, the chosen initial states imply that quantities involving unequal numbers of creation and annihilation operators vanish, e.g. $\braket{a_h}=\braket{a_h a_h}=0$. This simplifies the above calculation.

\subsection{Convergence}
\begin{figure}[t]
    \centering
\includegraphics[width=0.95\linewidth]{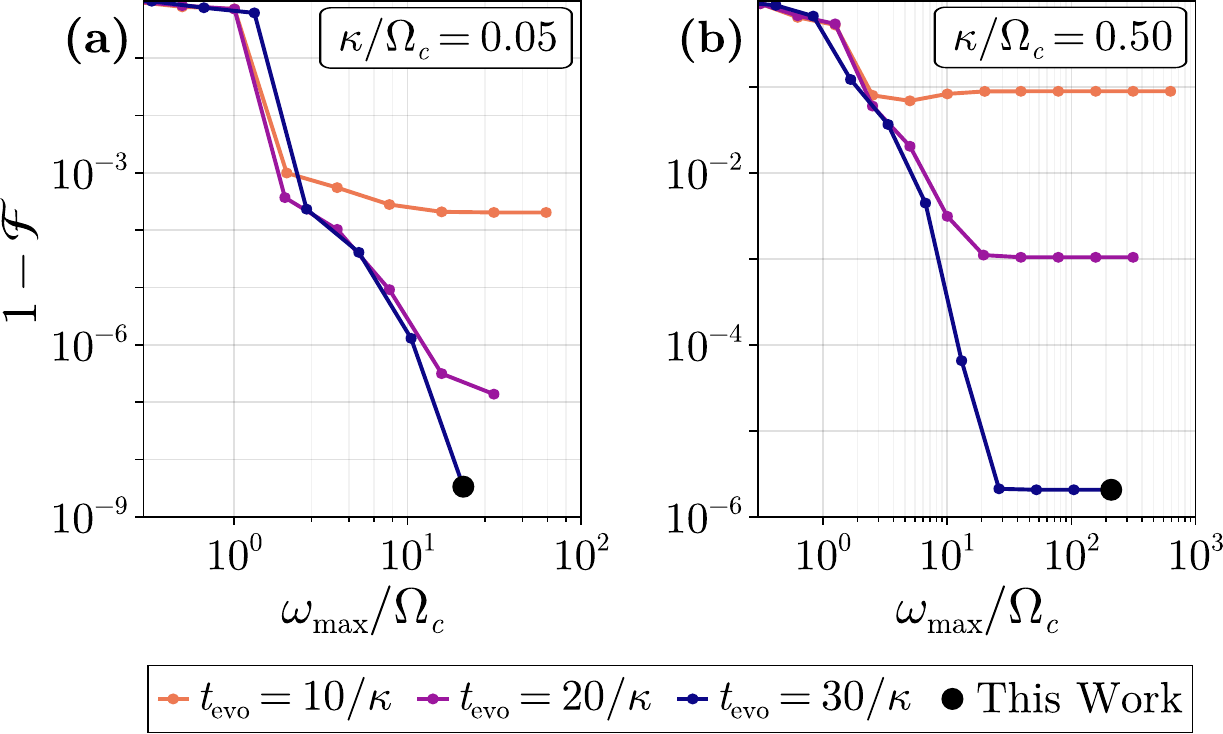}
    \caption{Infidelity $1-\mathcal{F}$ for the Exact method for various $t_{\mathrm{evo}}$ and $\omega_{\mathrm{max}}$ (corresponding to values of $N$ up to 2000) at weak and strong system-bath coupling $\kappa$. Fidelity is calculated against a reference obtained using $N_{\mathrm{bath}}=4000, t_{\mathrm{evo}}=40/\kappa$. 
    System parameters: $ T_h/\Omega_c=5, T_c/\Omega_c=0.5, \omega_\mathrm{cut}/\Omega_c=3.0, \Omega_h/\Omega_c=1.0, g/\Omega_c=0.5, \kappa_{h}=\kappa_c=\kappa$. Panel (a) $\kappa/\Omega_c=0.05$, panel (b) $\kappa/\Omega_c=0.50$. In this work, we use $N_{\mathrm{bath}}=2000, t_{\mathrm{evo}}=30/\kappa$, indicated by the black circle.} 
    \label{fig:exact:benchmark}
\end{figure}

The parameters used in this work, $N_{\mathrm{bath}}=2000, t_{\mathrm{evo}}=30/\kappa$, when compared against the reduced system density matrix obtained using $N_{\mathrm{bath}}=4000, t_{\mathrm{evo}}=40/\kappa$, produces a maximum infidelity of $1-\mathcal{F}<10^{-5}$ across all parameter regimes investigated. As show in Fig.~\ref{fig:exact:benchmark}, both the maximum bath frequency $\omega_{\mathrm{max}}=N_{\mathrm{bath}}\Delta \omega$ and $t_\mathrm{evo}$  determines the accuracy of the Exact method.

\section{HOPS}\label{app:HOPS}
\subsection{HOPS Equation of Motion}\label{app:HOPS:eom}
HOPS involves the evolution of physical and auxiliary state vectors whose evolutions are governed by coupled stochastic Schrödinger equations. The states are labelled by a vector $\vec{k}\in \mathbb{N}^{2 N_{\mathrm{BCF}}}$ of indices $k_{\ell\mu}$, $\ell = h,c$ and $\mu = 1,\dotsc,N_{\mathrm{BCF}}$.  $\vec{k} = \vec{0}$ encodes the physical state and $\vec{k} \neq \vec{0}$ the auxiliary states. $N_{\mathrm{BCF}}$ is the number of exponentials used to approximate the zero-temperature emission bath correlation function (Eq.~\eqref{bcf1}), 
\begin{equation}
    \bar{\alpha}_\ell^0(t) = \sum_{\mu = 1}^{N_{\mathrm{BCF}}} G_{\ell\mu} e^{-W_{\ell\mu}t},
    \label{app:HOPS:fitting}
\end{equation}
with coefficients $G_{\ell\mu}, W_{\ell\mu} \in \mathbb{C}$ which are fitted numerically with parameters in App.~\ref{app:HOPS:exp}.
The combined system and auxiliary objects can be elegantly described in an enlarged Hilbert space \cite{gao_non-markovian_2022, flannigan_many-body_2022} as
\begin{equation}\label{eq:defPhi}
    \Ket{\Phi(t)} = \sum_{\vec{k} \in \mathfrak{K}} |\psi^{(\vec{k})}(t)\rangle\ket{\vec{k}},
\end{equation}
\noindent
where each hierarchy index is represented by a bosonic mode and auxiliary Fock states $\ket{\vec{k}} = \otimes_{\ell = h,c} \otimes_{\mu = 1}^{N_{\mathrm{BCF}}} \ket{k_{\ell\mu}}$.
The associated bosonic operators $b_{\ell\mu}^\dagger, b_{\ell\mu}$ satisfy the standard relations
\begin{equation}
    \begin{split}
          b_{\ell\mu}^\dagger\ket{k_{\ell\mu}} & = \sqrt{k_{\ell\mu}+1} ~\ket{k_{\ell\mu}+1},\\
          b_{\ell\mu}\ket{k_{\ell\mu}} & = \sqrt{k_{\ell\mu}} ~\ket{k_{\ell\mu}-1}.
    \end{split}
\end{equation}
Note that in practice, a ``triangular" truncation rule is used: $\mathfrak{K} = \{ \vec{k}: ||\vec{k}||_1 \leq K\}$ \cite{zhang_flexible_2018}. \\

The method can work directly with a time-dependent Hamiltonian, and easily accepts modifications to the state vector as described in App.~\ref{app:HOPS:state}. The non-linear, un-normalised version of the equation of motion is as follows \cite{boettcher_dynamics_2024},
\begin{equation}
        \begin{aligned}
        \frac{\mathrm{d}}{\mathrm{d}t}&\ket{\Phi(t)} = \Big\{ -\i H_S^l(t) \Big.
                                    + \sum_{\ell = h,c} \left[ \tilde{\eta}_\ell^*(t) + \zeta_\ell^*(t) \right] L_\ell - \zeta_\ell(t) L_\ell^\dagger\\
                                    & - \sum_{\ell = h,c} \sum_{\mu=1}^{N_{\mathrm{BCF}}} W_{\ell\mu} b_{\ell\mu}^\dagger b_{\ell\mu}  -\i \sum_{\ell = h,c} \sum_{\mu=1}^{N_{\mathrm{BCF}}} \sqrt{G_{\ell\mu}}  L_\ell b_{\ell\mu}^\dagger \\
                                    & \Big. -\i \sum_{\ell = h,c} \sum_{\mu=1}^{N_{\mathrm{BCF}}} \sqrt{G_{\ell\mu}} (L_\ell^\dagger - \braket{L_\ell^\dagger}_t) b_{\ell\mu} \Big\}\ket{\Phi(t)},
    \end{aligned}
    \label{eq:HOPS}
\end{equation}
where $\tilde{\eta}_\ell^*(t)$ is a shifted coloured noise for each bath $\ell = h,c$, defined as follows \cite{hartmann_open_2021},
\begin{equation}
\begin{split}
    &\tilde{\eta}_\ell^*(t)=\eta_\ell^*(t)+\eta_{\mathrm{sh},\ell}^*(t),\\
    &\eta_{\mathrm{sh},\ell}^*(t) =  \int_0^t \bar{\alpha}_\ell^{0*}(t-s)\braket{L^\dagger}_s ~ \mathrm{d} s,
\end{split}
\end{equation}
\noindent
with the coloured noise $\eta_\ell^*(t)$ satisfying the  properties
\begin{equation}
\begin{split}
    &\mathcal{M}\{\eta_\ell(t)\} = \mathcal{M}\{\eta_\ell(t) \eta_\ell(s)\} = 0,\\
    &\mathcal{M}\left\{\eta_\ell(t)\eta_\ell(s)^*\right\} = \bar{\alpha}^0_\ell(t-s).
\end{split}
    \label{eq:HOPS:z}
\end{equation}
Using the exponential fitting of $\bar{\alpha}_\ell^{0*}(t-s)$, $\eta_{\mathrm{sh},\ell}^*(t)$ evolves as follows \cite{hartmann_open_2021}, 
\begin{equation}
    \frac{\mathrm{d}}{\mathrm{d}t}\eta_{\mathrm{sh},\ell}^*(t) = \sum_{\mu=1}^{N_{\mathrm{BCF}}} G_{\ell\mu}^* \Braket{L_\ell^\dagger}_t-W_{\ell\mu}^* \eta_{\mathrm{sh},\ell}^*(t).
    \label{eq:HOPS:noise_shift}
\end{equation}

The collection of hierarchy states and the noise shift comprise all the information about the system and baths, and are evolved time-locally using RK4.
 
We emphasize that the physical trajectory state $|\psi^{(\vec{0})}(t)\rangle$ is obtained from the solution of Eq.~\eqref{eq:HOPS} by projecting on the vacuum state of the hierarchy bosonic modes $|\psi^{(\vec{0})}(t)\rangle = \langle0|\Phi(t)\rangle$, according to Eq.~\eqref{eq:defPhi}.
For the numerical solution, we perform a soft normalisation by removing the component of the infinitesimal change $\Ket{\rm{d}\Phi(t)}$ along the physical state vector $|\psi^{(\vec{0})}(t)\rangle$,
\begin{equation}
    \Ket{\rm{d}\Phi(t)} \to \Ket{\rm{d}\Phi(t)} - \Braket{\rm{d}\psi^{(\vec{0})}(t)|\psi^{(\vec{0})}(t)}\Ket{\Phi(t)}/||\psi^{(\vec{0})}(t)||^2.
\end{equation}

After each time step, we normalise the state again, dividing by $||\psi^{(\vec{0})}(t)||$. During each trajectory, observables are calculated using the normalised physical state vector $|\tilde{\psi}^{(\vec{0})}(t)\rangle = |\psi^{(\vec{0})}(t)\rangle/||\psi^{(\vec{0})}(t)||$, for example, 
\begin{equation}
    \braket{L^\dagger}_t = \Bra{\tilde{\psi}^{(\vec{0})}(t)}L^\dagger \Ket{\tilde{\psi}^{(\vec{0})}(t)}.
\end{equation}
This is also denoted as $ \mathbb{E}[L^\dagger] $.

In this method, the effect of temperature is implemented via a second coloured noise $\zeta_\ell(t)$, satisfying \cite{hartmann_open_2021}
\begin{equation}
\begin{split}
    &\mathcal{M}\{\zeta_\ell(t)\} = \mathcal{M}\{\zeta_\ell(t) \zeta_\ell(s)\} = 0,\\
    &\mathcal{M}\{\zeta_\ell(t)\zeta_\ell(s)^*\}= \int_0^\infty n_\ell(\omega) J_\ell(\omega) e^{-\i \omega (t-s)} ~\mathrm{d}\omega,
\end{split}
\label{eq:HOPS:xi}
\end{equation}
\noindent
while keeping the bath correlation function $\bar{\alpha}_\ell^0(t)$ at zero temperature. The generation of noises $\eta_\ell(t), \zeta_\ell(t)$ is described in App.~\ref{app:HOPS:noise}.\\

The reduced density matrix for the junction is obtained by averaging the density matrix corresponding to the normalised state vector over many trajectories, each of which is based on an i.i.d. realisation of the two noises, 
\begin{equation}
    \rho(t) = \mathcal{M}\left\{\Ket{\tilde{\psi}^{(\vec{0})}(t)}\Bra{\tilde{\psi}^{(\vec{0})}(t)}\right\}.
    \label{eq:HOPS:rho}
\end{equation}
\subsection{Modifications to State Vector} \label{app:HOPS:state}
\subsubsection{Displacement Method}
It was found that the number of Fock states for the two system harmonic oscillators can be drastically reduced by compressing the occupation of the higher levels. After each time-step, $\alpha_\ell=\braket{a_\ell}$ is evaluated, then a displacement operator $D_\ell(-\alpha_\ell)$
is applied to the system, where $D_\ell(\beta)=\exp(\beta a_\ell^\dagger -\beta^* a_\ell)$. 
Repeated applications of displacement operators only incur a global phase, which would cancel out when calculating expectation values.
In the displaced frame, all the operators in the Hamiltonian will need to be conjugated accordingly, 
\begin{equation}
    a_\ell^{\mathrm{hf}} =D_\ell(-\braket{a_\ell}) a_\ell D_\ell(-\braket{a_\ell})^\dagger =a_\ell + \braket{a_\ell}.
\end{equation}
This allows us to use very few Fock states to represent the system bosonic sites (App.~\ref{app:HOPS:conv}).

\subsubsection{On-site Interactions}
The introduction of non-linear on-site interactions causes the system to become stiff and difficult to evolve using standard methods. 
To deal with this, we choose an interaction picture with respect to the following, 
\begin{equation}
    H_0 = \sum_{\ell = L,R} \Omega_\ell n_\ell + \frac{U}{2} n_\ell (n_\ell-\mathds{1}),
\end{equation}
\noindent
where $n_\ell =a_\ell^\dagger a_\ell$. Then
the lowering and raising operators become
\begin{equation}
    \begin{split}
        a_\ell^{\mathrm{hf}} & = e^{\i H_0t}a_\ell e^{-\i H_0t} \\
        & = \sum_{n=0}^\infty e^{-\i[\Omega_\ell + U (n-1)]t} \sqrt{n} \ket{n-1}\bra{n}, \\
        {a_\ell^{\mathrm{hf}}}^\dagger & = \sum_{n=0}^\infty e^{\i[\Omega_\ell + U (n-1)]t} \sqrt{n} \ket{n}\bra{n-1}.
    \end{split} 
\end{equation}
This change can be made consistently to all operators. This resolves the stiffness issue.

\subsection{Analytic Expressions for System-Bath Observables}\label{app:HOPS:obs}
To derive expectation values involving bath observables, we need to follow through the HOPS formalism and apply all the transformations in order.

\subsubsection{HOPS Formalism} \label{app:HOPS:obs:formalism}
As in App.~\ref{app:HOPS:noise}, for each trajectory we have random complex samples $\mathbf{z}_\ell=(z_{\ell1}, z_{\ell2}, ...)$ for coloured noise $\eta_\ell(t)$, 
and $\mathbf{y}_\ell = (y_{\ell1}, y_{\ell2}, ...)$ for temperature noise $\zeta_\ell(t)$.

We make the transformation into the rotating frame with respect to the free bath Hamiltonian (Eq.~\eqref{eq:Hamiltonian_lab_3}), the resulting system and bath evolves with Hamiltonian
\begin{equation}
H^{\mathrm{HOPS}}=H_S^{\mathrm{lf}}(t)+\sum_{\ell,k}\left(g_{\ell k} a_\ell^\dagger a_{\ell k} e^{-\i \omega_{\ell k}t}+\mathrm{h.c.}\right).
\end{equation}

The system and bath starts in state $\rho_{\rm{sys + bath}}(0) = \rho_\mathrm{sys}(0) \otimes  \left(\bigotimes_{\ell=h,c} \rho^\mathrm{thermal}_\ell\right)$.
Let coherent states $\ket{\mathbf{y}} = D(\mathbf{y}) \ket{\mathbf{0}}$,
where $D(\mathbf{y})=\prod_{\ell,k} \exp(y_{\ell k} a_{\ell k}^\dagger -y_{\ell k} ^* a_{\ell k} )$ is the (normalised) displacement operator for a bosonic bath. Here, $k$ is the index of the bath modes. \\

The thermal density matrix is given by the following,
where we define the abbreviation $\mathcal{M}_{\mathbf{y}}$ to represent the integral as the average over random samples,
\begin{equation}
    \begin{split}
        \rho^{\rm{thermal}} & = \bigotimes_{\ell, k} \int d^2 y_{\ell k} \frac{1}{\pi n(\beta, \omega_{\ell k})} e^{-|y_{\ell k}|^2/n(\beta, \omega_{\ell k})} \ket{y_{\ell k}}\bra{y_{\ell k}}\\
                        & := \mathcal{M}_{\mathbf{y}} \{\ket{\mathbf{y}}\bra{\mathbf{y}}\}\\
                        & = \mathcal{M}_{\mathbf{y}} \{D(\mathbf{y}) \ket{\mathbf{0}}\bra{\mathbf{0}}D(\mathbf{y})^\dagger\}.
    \end{split}
\end{equation}
\noindent
Let $U(t)$ be the unitary operator that evolves the combined system and bath in this rotating frame, then we have
\begin{equation}
    \begin{split}
        \rho_{\rm{sys+bath}}(t) & = U(t) \rho_{\rm{sys+bath}}(0) U(t)^\dagger \\
                                & = \mathcal{M}_{\mathbf{y}} \{U(t) ~ \rho_\mathrm{sys}(0) (D(\mathbf{y}) \ket{\mathbf{0}}\bra{\mathbf{0}}D(\mathbf{y})^\dagger) ~ U(t)^\dagger\}.
    \end{split}
\end{equation}

Define $U(\mathbf{y}, t) = D(\mathbf{y})^\dagger U(t) D(\mathbf{y})$, this allows us to write 
\begin{equation}
    \begin{split}
        \rho_{\rm{sys+bath}}(t) & = \mathcal{M}_{\mathbf{y}} \{D(\mathbf{y}) U(\mathbf{y}, t) ~ \rho_\mathrm{sys}(0) \ket{\mathbf{0}}\bra{\mathbf{0}} ~ U(\mathbf{y}, t)^\dagger D(\mathbf{y})^\dagger \}.
    \end{split}
\end{equation}

We set the system to be initialised as a pure state,
$\rho_\mathrm{sys}(0) = \ket{\psi(0)}\bra{\psi(0)}$.
Then define $\ket{\Psi(\mathbf{y}, t)} = \tilde{U}(\mathbf{y}, t) \ket{\psi(0)} \otimes \ket{\mathbf{0}}$ as the evolution of the system and bath.
This is equivalent to evolving under the shifted hamiltonian $\tilde{H}^{\mathrm{HOPS}}(\mathbf{y}) = D(\mathbf{y})^\dagger H^{\mathrm{HOPS}} D(\mathbf{y})$,
which is the origin of the temperature noise terms in Eq.~\eqref{eq:HOPS}.

At this point, we apply other possible transformations of the system (App.~\ref{app:HOPS:state}). We denote the resulting frame of the above operations as the 'HOPS frame', indicated by the superscript $\mathrm{hf}$. The bath operators in this frame are
\begin{equation}
    \begin{split}
        a_{\ell k}^{\mathrm{hf}} & = (a_{\ell k} + y_{\ell k}) e^{-\i \omega_{\ell k} t}, \\
        {a_{\ell k}^{\mathrm{hf}}}^\dagger & = (a_{\ell k}^\dagger + y_{\ell k}^*) e^{\i \omega_{\ell k} t}.
    \end{split}
\end{equation}
\subsubsection{Bath Occupancy}\label{app:HOPS:bath_occupancy}
Here we demonstrate the frame transformations explicitly. The occupancy of the bath mode with frequency $\omega_{\ell k}$ is
\begin{equation}
    \begin{split}
        \braket{a_{\ell k}^\dagger a_{\ell k}} & = {\rm Tr}_{\rm{sys + bath}} [\rho_{\rm{sys + bath}} a_{\ell k}^\dagger a_{\ell k}]\\
                                             & = \mathcal{M}_{\mathbf{y}} \left\{{\rm Tr}\left[D(\mathbf{y}) ~ \ket{\Psi(\mathbf{y}, t)}\bra{\Psi(\mathbf{y}, t)} ~ D(\mathbf{y})^\dagger ~ a_{\ell k}^\dagger a_{\ell k}\right]\right\}\\
                                             & = \mathcal{M}_{\mathbf{y}}  \left\{\bra{\Psi(\mathbf{y}, t)} D(\mathbf{y})^\dagger a_{\ell k}^\dagger a_{\ell k}  D(\mathbf{y}) \ket{\Psi(\mathbf{y}, t)}\right\}.\\
    \end{split}
\end{equation}
We may use coherent state identities and commutation relations to evaluate
\begin{equation}
    \begin{split}
        D(\mathbf{y})^\dagger a_{\ell k}^\dagger a_{\ell k} D(\mathbf{y}) & = (a_{\ell k}^\dagger + y_{\ell k}^*)(a_{\ell k} + y_{\ell k})\\
                                                             & = a_{\ell k}^\dagger a_{\ell k} + y_{\ell k}^* a_{\ell k} + y_{\ell k} a_{\ell k}^\dagger + |y_{\ell k}|^2\\
                                                             & = a_{\ell k} a_{\ell k}^\dagger - \mathds{1} + y_{\ell k}^* a_{\ell k} + y_{\ell k} a_{\ell k}^\dagger + |y_{\ell k}|^2.
    \end{split}
\end{equation}
We have the identity operator 
\begin{equation}
    \mathds{1} = \bigotimes_k \int \mathrm{d}^2 z_{\ell k} ~\frac{1}{\pi} e^{-|z_{\ell k}|^2} \ket{z_{\ell k}}\bra{z_{\ell k}},
\end{equation}
where $\ket{z_{\ell k}}= \exp(z_{\ell k} a_{\ell k}^\dagger)\ket{0}$ is the un-normalised coherent state. Inserting this \textbf{in-between} $a_{\ell k} a_{\ell k}^\dagger$ allows us to use the identity
$a_{\ell k} \ket{z_{\ell k}} = z_{\ell k} \ket{z_{\ell k}}$. We also abbreviate the integral as $\mathcal{M}_{\mathbf{z}}$.

Define $\ket{\psi(\mathbf{z}, \mathbf{y}, t)}:= \braket{\mathbf{z}| \Psi(\mathbf{y}, t)}$. Then we have
\begin{equation}
    \begin{split}
        \braket{a_{\ell k}^\dagger a_{\ell k}} 
                                                & = \mathcal{M}_{\mathbf{y},\mathbf{z}} \left\{ \braket{\psi(\mathbf{z}, \mathbf{y}, t)|\psi(\mathbf{z}, \mathbf{y}, t)} \left(|z_{\ell k} + y_{\ell k} |^2 -1\right)\right\}.
    \end{split}
\end{equation}

The Girsanov transform \cite{diosi_non-markovian_1998} allows us to replace the expectation over unnormalised state vectors with expectation over normalised state vectors evolved using a shifted noise $\tilde{z}_k(t)$.
\begin{equation}
    \begin{split}
        \mathcal{M}_{\mathbf{y},\mathbf{z}} & \left\{ \ket{\psi(\mathbf{z}, \mathbf{y}, t)}\bra{\psi(\mathbf{z}, \mathbf{y}, t)}  \right\} \\ &= \mathcal{M}_{\mathbf{y},\mathbf{z}} \left\{ \ket{\tilde{\psi}(\mathbf{\tilde{z}}(t), \mathbf{y}, t)}\bra{\tilde{\psi}(\mathbf{\tilde{z}}(t), \mathbf{y}, t)}  \right\}.
    \end{split}
\end{equation}
Dropping the subscripts for $\mathcal{M}$ for clarity, we have
\begin{equation}
    \begin{split}
        \braket{a_{\ell k}^\dagger a_{\ell k}} & = \mathcal{M} \left\{\left(|\tilde{z}_{\ell k}(t) + y_{\ell k} |^2 -1\right)\right\}.
    \end{split}
\end{equation}
The shifted noise elements are given by the following:
\begin{equation}
    \begin{split}
        \tilde{z}_{\ell k}^*(t) = z_{\ell k}^* + \i g_{\ell k} \int_0^t ds ~ e^{-\i \omega_{\ell k} s} \braket{L^\dagger}_s,
    \end{split}
\end{equation}
which we can compute from the value of $\braket{L^\dagger}$ at each time-step. This calculation is not intensive if we only do it for those $\omega_{\ell k}$ within plotting range. These relate to the shifted noise by
\begin{equation}
    \begin{split}
        \tilde{\eta}_\ell^*(t) = -\i \sum_{k = 1}^N g_{\ell k}^* \tilde{z}_{\ell k}^* e^{\i \omega_{\ell k} t}.
    \end{split}
\end{equation}
Including the bath subscripts $\ell=h,c$ gives the bath mode occupation for the trajectory
\begin{equation}
    \begin{split}
        \braket{a_{\ell k}^\dagger a_{\ell k}}(\tilde{z}_{\ell k}(t),y_{\ell k} , t) & = \mathcal{M} \left\{ |\tilde{z}_{\ell k}(t) + y_{\ell k} |^2 -1\right\}.
    \end{split}
\end{equation}

\subsubsection{Momentum-resolved Particle Currents}\label{app:HOPS:gradient}
In order to obtain the gradient of the average bath mode occupation $\dot{\langle a_{k\ell}^\dagger a_{k\ell}\rangle}$, we perform an Ordinary Least Squares (OLS) regression on the converged section of the time series. We are able to reduce memory requirements by noting that the gradient from OLS is a linear transformation on input data \cite{rao_linear_1973}. Therefore we can first perform the regression at the trajectory level, then average over trajectories to get
\begin{equation}
    \dot{\langle a_{k\ell}^\dagger a_{k\ell}\rangle} = \frac{\mathrm{d}}{\mathrm{d}t}\mathcal{M}\left\{\braket{a_{\ell k}^\dagger a_{\ell k}}\right\} = \mathcal{M}\left\{\frac{\mathrm{d}}{\mathrm{d}t}\braket{a_{\ell k}^\dagger a_{\ell k}}\right\}.
\end{equation}
Note that this does not hold for other types of regression. 

\subsubsection{Analytic Power and Heat Currents}
Similarly, we obtain the following when using the above formalism to evaluate Eq.~\eqref{def:currents_exact_lab}, resulting in
\begin{align}
        \mathcal{P} & =  \mathcal{M}\left\{-2 g\mathcal{E}\Im\left[ \mathbb{E}\left({a_h^{\mathrm{hf}}}^\dagger {a_c^{\mathrm{hf}}}\right)\right]\right\},\label{def:power_HOPS_analytical}\\
    \mathcal{J}_\ell & =  \mathcal{M}\left\{ -2\Re\left[\left(\tilde{\eta}_\ell^*(t) + \zeta_\ell^*(t)\right)\mathbb{E}\left( S_\ell^{\mathrm{hf}}\right) \right] \right\},\label{def:heat_currents_HOPS_analytical}
\end{align}
where
\begin{equation}
    S_\ell^{\mathrm{hf}} = \begin{cases}
        \Omega_\ell a_\ell^{\mathrm{hf}} + g a_{\bar{\ell}}^{\mathrm{hf}} +U n_\ell^{\mathrm{hf}} a_\ell^{\mathrm{hf}} & \text{if } U\geq 0\\
        \Omega_\ell a_\ell^{\mathrm{hf}} + g (\mathds{1}-2n_\ell^{\mathrm{hf}}) a_{\bar{\ell}}^{\mathrm{hf}} & \text{if } U\to \infty\\
    \end{cases},
\end{equation}

In App.~\ref{app:HOPS:comparison} we show that the two expressions produce the same steady-state predictions. HOPS offers one a choice when deriving formally equivalent expressions to utilise either the coloured noises $\tilde{\eta}_\ell(t), \zeta_\ell(t)$ or the auxiliary state vectors $\ket{\tilde{\psi}^{(\vec{k})}(t)}$ \cite{boettcher_dynamics_2024}. The empirical expression implicitly involves both. Further investigation is required to understand the impact of the hierarchy truncation $k$ on the accuracy and variance of different expressions.

\subsection{Empirical Power and Heat Current} \label{app:HOPS:heat_current}
The equation of motion is evolved using the RK4 scheme.
For brevity, define the functions $f_\ell$ which compose Eq.~\eqref{eq:HOPS} as follows,
\begin{equation}
    \begin{split}
        \frac{\mathrm{d}}{\mathrm{d}t}\ket{\Phi(t)} & := -\i H_S^l \ket{\Phi(t)} + \sum_{\ell=h,c} f_\ell\left(\ket{\Phi(t)}, \eta_{\mathrm{sh}}^*(t), t\right). \\
    \end{split}
\end{equation}

Explicitly, this is 
\begin{equation}
        \begin{split}
        f_\ell\left(\ket{\Phi(t)},\right.&\left. \eta_{\mathrm{sh}}^*(t), t\right)  = \left\{\left[ \eta_\ell^*(t) + \eta_{\mathrm{sh}}^*(t)+ \zeta_\ell^*(t) \right] L_\ell \right.\\
                                    & - \zeta_\ell(t) L_\ell^\dagger- \sum_{\mu=1}^{N_{\mathrm{BCF}}} W_{\ell\mu} b_{\ell\mu}^\dagger b_{\ell\mu} \\
                                    & -\i \sum_{\mu=1}^{N_{\mathrm{BCF}}} \sqrt{G_{\ell\mu}}  L_\ell b_{\ell\mu}^\dagger \\
                                    & \left. -\i  \sum_{\mu=1}^{N_{\mathrm{BCF}}} \sqrt{G_{\ell\mu}} (L_\ell^\dagger - \braket{L_\ell^\dagger}_t) b_{\ell\mu} \right\}\ket{\Phi(t)}.
    \end{split}
    \label{app:eq:HOPS_components}
\end{equation}

During each time-step, let the RK4 increments of $\Ket{\Phi(t)}$ be $\ket{\phi_1}, \ket{\phi_2}, \ket{\phi_3}, \ket{\phi_4}$, and of the noise shift (Eq.~\eqref{eq:HOPS:noise_shift}) be $\eta_{\mathrm{sh}1}, \eta_{\mathrm{sh}2}, \eta_{\mathrm{sh}3}, \eta_{\mathrm{sh}4}$. The timestep update is given by

\begin{equation}
    \begin{split}
        \Ket{\Phi(t+\Delta t)}&:=\Ket{\Phi(t)} + \frac{\Delta t}{6}\left(\ket{\phi_1}+2 \ket{\phi_2}+2 \ket{\phi_3}+ \ket{\phi_4}\right),\\
        \eta_{\mathrm{sh}}(t+\Delta t)&:=\eta_{\mathrm{sh}}(t)+\frac{\Delta t}{6}\left(\eta_{\mathrm{sh}1}+2\eta_{\mathrm{sh}2}+2\eta_{\mathrm{sh}3}+\eta_{\mathrm{sh}4}\right).
    \end{split}
\end{equation}

Correspondingly, let the components corresponding to each bath be 
\begin{align}
        \ket{\phi_{1,\ell}} & = f_\ell\left(\ket{\Phi(t)}, \eta_{\mathrm{sh}}^*(t), t\right), \\
        \ket{\phi_{2,\ell}} & = f_\ell\left(\ket{\Phi(t)} + \frac{\Delta t}{2} \ket{\phi_1}, \eta_{\mathrm{sh}}^*(t) + \frac{\Delta t}{2} \eta_{\mathrm{sh}1}^*, t + \frac{\Delta t}{2}\right), \\
        \ket{\phi_{3,\ell}} & = f_\ell\left(\ket{\Phi(t)} + \frac{\Delta t}{2} \ket{\phi_2}, \eta_{\mathrm{sh}}^*(t) + \frac{\Delta t}{2} \eta_{\mathrm{sh}2}^*, t + \frac{\Delta t}{2}\right), \\
        \ket{\phi_{4,\ell}} & = f_\ell\left(\ket{\Phi(t)} + \Delta t \ket{\phi_3}, \eta_{\mathrm{sh}}^*(t) + \Delta t \eta_{\mathrm{sh}3}^*, t + \Delta t\right). 
    \end{align}
\noindent
Denote the intermediate states (which contains both system and auxiliary levels) as used during the RK4 computation to advance from time $t$ to time $t+\Delta t$ to be
\begin{equation}
    \begin{split}
        \Ket{\Phi_1}&:=\ket{\Phi(t)} + \frac{\Delta t}{2} \ket{\phi_1},\\
        \Ket{\Phi_2}&:=\ket{\Phi(t)} + \frac{\Delta t}{2} \ket{\phi_2},\\
        \Ket{\Phi_3}&:=\ket{\Phi(t)} + \Delta t \ket{\phi_3}.\\
    \end{split}
\end{equation}

We can then extract the \textit{unnormalised} system state vector by taking the component corresponding to zero auxiliary excitations,
\begin{equation}
    \begin{split}
        \ket{\psi^{(\vec{0})}_0}&:=\braket{\vec{0}|\Phi(t)},\\
        \ket{\psi^{(\vec{0})}_1}&:=\braket{\vec{0}|\Phi_1},\\
        \ket{\psi^{(\vec{0})}_2}&:=\braket{\vec{0}|\Phi_2},\\
        \ket{\psi^{(\vec{0})}_3}&:=\braket{\vec{0}|\Phi_3}.\\
    \end{split}
\end{equation}

The fourth-order estimate for the heat current (Eq.~\ref{def:HOPS:currents}) averaged over a time-step similarly uses the weights and  intermediate stages \cite{butcher_numerical_2016}, giving
\begin{equation}
    \begin{split}
        \mathcal{J}_\ell = 2 \cdot \Re & \left\{ \frac{1}{6}\left[ 
            \Braket{\psi^{(\vec{0})}_0 | H_S^l(t) | \phi^{(\vec{0})}_{1,\ell}} /|| \psi^{(\vec{0})}_0||^2 \right. \right.\\
            & + 2\Braket{\psi^{(\vec{0})}_1  | H_S^l\left(t + \frac{\Delta t}{2}\right) | \phi^{(\vec{0})}_{2,\ell}}/|| \psi^{(\vec{0})}_1||^2\\
            & + 2\Braket{\psi^{(\vec{0})}_2 | H_S^l\left(t + \frac{\Delta t}{2}\right) | \phi^{(\vec{0})}_{3,\ell}}/|| \psi^{(\vec{0})}_2||^2\\
            & + \left. \left. \Braket{\psi^{(\vec{0})}_3  | H_S^l\left(t + \Delta t\right) | \phi^{(\vec{0})}_{4,\ell}}/|| \psi^{(\vec{0})}_3||^2 \right] \right\}.
    \end{split}
    \label{eq:HOPS:heat_current_empirical}
\end{equation}
\noindent
The fourth-order estimate for the power $\mathcal{P}$ (Eq.~\ref{def:HOPS:power}) averaged over a time-step is
\begin{equation}
    \begin{split}
        \mathcal{P} =-& \frac{1}{6}\left[ 
            \Braket{\psi^{(\vec{0})}_0 | \dot{H_S^l}(t) | \psi_0^{(\vec{0})}} \right./|| \psi_0^{(\vec{0})}||^2 \\
            & + 2\Braket{\psi^{(\vec{0})}_1 | \dot{H_S^l}\left(t + \frac{\Delta t}{2}\right) | \psi^{(\vec{0})}_1  }/|| \psi^{(\vec{0})}_1||^2\\
            & + 2\Braket{\psi^{(\vec{0})}_2 | \dot{H_S^l}\left(t + \frac{\Delta t}{2}\right) | \psi^{(\vec{0})}_2 }/|| \psi^{(\vec{0})}_2||^2\\
            & +  \left. \Braket{\psi^{(\vec{0})}_3 |\dot{H_S^l}\left(t + \Delta t\right) |\psi^{(\vec{0})}_3 } /|| \psi^{(\vec{0})}_3||^2\right],
    \end{split}
    \label{eq:HOPS:power_empirical}
\end{equation}
\noindent where
\begin{align}
    \dot{H_S^l}(t) &=\i g\mathcal{E}\left( a_c^\dagger a_h e^{\i\mathcal{E}t}-a_c a_h^\dagger e^{-\i\mathcal{E}t}\right).
\end{align}

\subsection{Comparison of HOPS Empirical and Analytical Expressions}\label{app:HOPS:comparison}
We observe excellent agreement in the steady-state value of power and heat currents obtained using the empirical (Eqs.~\eqref{eq:HOPS:power_empirical}, \eqref{eq:HOPS:heat_current_empirical}) and analytical (Eqs.~\eqref{def:power_HOPS_analytical}, \eqref{def:heat_currents_HOPS_analytical}) expressions. Note that observables are plotted every 100 time-steps to reduce memory requirements.
\begin{figure}[t]
    \centering
    \includegraphics[width=0.95\linewidth]{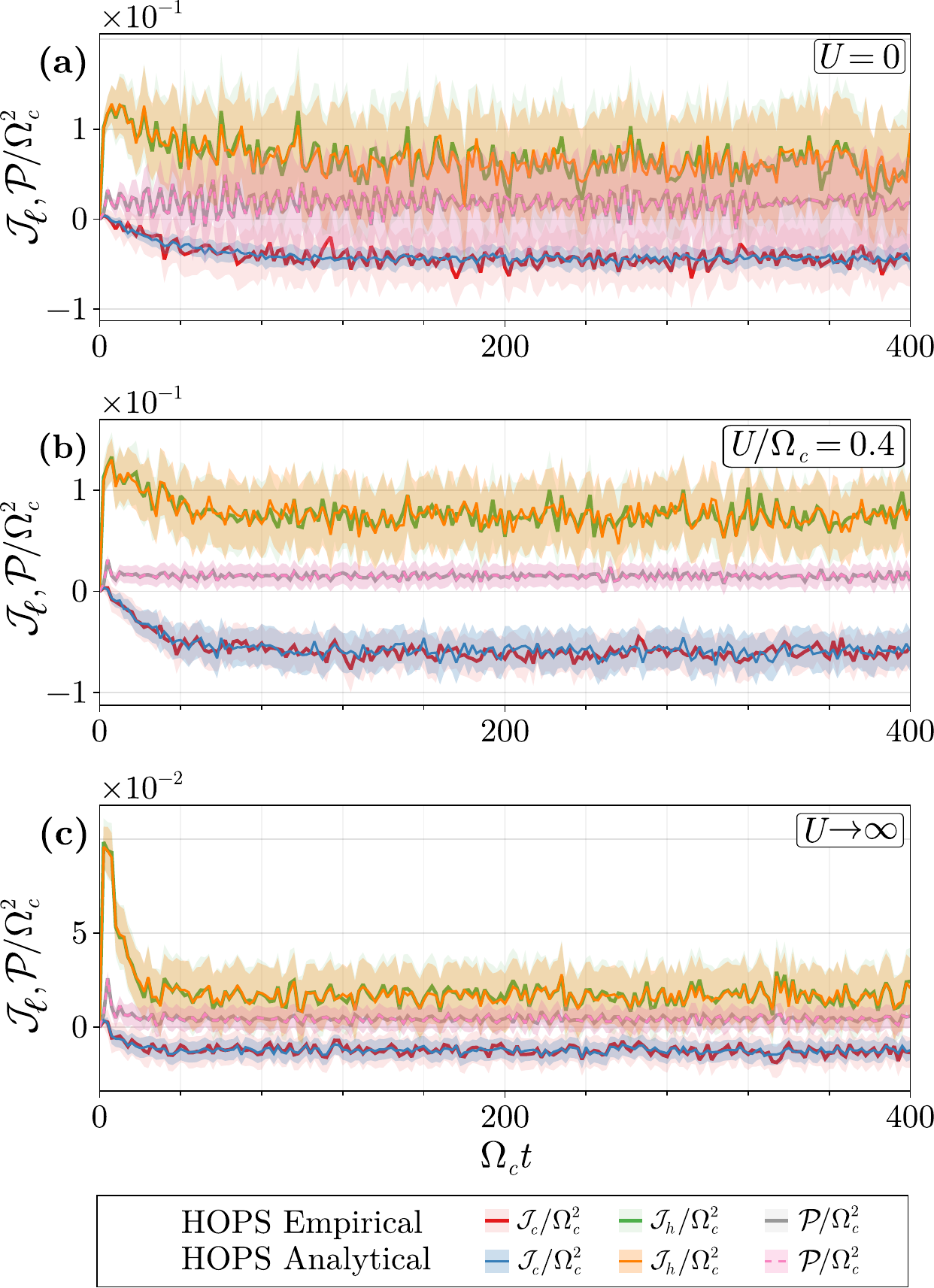}
    \caption{Comparison of HOPS Empirical (Eqs.~\eqref{eq:HOPS:power_empirical}, \eqref{eq:HOPS:heat_current_empirical}) and Analytical (Eqs.~\eqref{def:power_HOPS_analytical}, \eqref{def:heat_currents_HOPS_analytical}) expressions for evaluating the power $\mathcal{P}$ and heat currents $\mathcal{J}_\ell$ for $\ell=h,c$. System Parameters: $\Omega_h/\Omega_c=1.4, g/\Omega_c=0.5, T_h/\Omega_c=5.0, T_c/\Omega_c=0.5, \kappa_{h,c}/\Omega_c=0.05, \omega_{\mathrm{cut}}/\Omega_c=3.0$, for (a) harmonic oscillators $U=0$, (b) anharmonic oscillators $U/\Omega_c=0.4$, (c) two qubits $U\to \infty$. HOPS parameters are listed in Table \ref{table:HOPS:params}.}
    \label{app:fig:HOPS_empirical_analytic}
\end{figure}
The amplitude of the oscillation converges to zero as the number of trajectories increases to infinity. In practice, with 1000 trajectories, the amplitude is considerable. We evolve the system for a sufficiently long time such that system observables have converged. Fig.~\ref{app:fig:HOPS_empirical_analytic} shows that the time required for observables to reach steady-state decreases with interaction strength $U$. The mean over the section of the converged time series is then taken as an estimate of the steady-state value predicted by HOPS under the ergodic assumption.

\subsection{Fitting Exponentials to Bath Correlation Function} \label{app:HOPS:exp}
To find the coefficients $G_{\ell\mu}, W_{\ell\mu} \in \mathbb{C}$ in Eq.~\eqref{app:HOPS:fitting}, 
we used the package \texttt{BCFUtils} \cite{hartmann_2025_bcfutils}
with settings listed in Table \ref{app:HOPS:params_table}.

\begin{table}[t]
\centering
    \begin{tabular}{ |c|c|p{5cm}| } 
        \hline
         Parameter & Value & Usage \\ 
         \hline
         $s$ & $1$ & Exponent in Ohmic spectral density $J_\ell (\omega) \propto \omega^s e^{-\omega/\omega_\mathrm{cut}}$. \\ 
          \hline
         $\omega_\mathrm{cut}$ & 3 & Exponential cut-off frequency in $J_\ell$. \\ 
          \hline
         $t_{max}$ & $400/\Omega_\ell$ & Maximum time to fit until. \\ 
         \hline
          diff kind & abs p diff & Loss function is the absolute $p$-norm difference. \\ 
           \hline
          $p$ & $5$ & $p$-norm. \\ 
         \hline
          max iters & $1000$ & Maximum iterations. \\ 
         \hline
           num samples & $100$ & Number of randomly sampled initial conditions. \\ 
         \hline
\end{tabular}
\caption{Parameters for \texttt{BCFUtils}}
\label{app:HOPS:params_table}
\end{table}

\subsection{Description of Coloured Noise} \label{app:HOPS:noise}
The coloured noises $\eta_\ell(t), \zeta_\ell(t)$ are defined as follows to achieve the desired correlation functions (Eqs.~\eqref{eq:HOPS:z}, \eqref{eq:HOPS:xi}),
\begin{equation}
    \begin{split}
        \eta_\ell^*(t) & := -\i \sum_{k = 1}^N g_{\ell k}^* z_{\ell k}^* e^{\i \omega_{\ell k} t}, \\
        \zeta_\ell^*(t) & := -\i \sum_{k = 1}^N  g_{\ell k}^* y_{\ell k}^* e^{\i \omega_{\ell k} t}, \\
        y_{\ell k}& := x_{\ell k}\sqrt{n_\ell (\omega_{\ell k})} 
    \end{split}
\end{equation}
where $z_{\ell k}, x_{\ell k} \sim \mathcal{CN}(0,1)$ are i.i.d. complex normal random variables, and $n_\ell(\omega)$ is the Bose-Einstein distribution for bath $\ell$ with temperature $T_\ell$. The sum corresponds to a discretisation of the integral in Eq.~\eqref{eq:HOPS:xi}, with 
$\omega_{\ell k} = \omega_0 + k \cdot\Delta \omega$. In order for the sum to be computed using Fast Fourier Transform (FFT), we choose $\Delta \omega = \frac{2 \pi}{N_{\mathrm{step}} \cdot \delta t}$, where $N_{\mathrm{step}}$ is the number of time-steps, each of duration $\delta t$, with $N_{\mathrm{step}}\delta t=t_{\mathrm{evo}}$. We also choose $\omega_0 = \delta \omega /2$ to avoid singularities at $\omega=0$. Then for the time-steps $j=0,1...,N_{\mathrm{step}}-1$, we have

\begin{equation}
    \begin{split}
        \eta_\ell^*(j \cdot \delta t) & := -\i e^{\i \omega_0 j \delta t} \mathcal{FFT}\left(\sqrt{J_\ell(\omega_{\ell k}) \Delta \omega} ~ z_{\ell k}^* \right),\\
        \zeta_\ell^*(j \cdot \delta t) & := -\i e^{\i \omega_0 j \delta t} \mathcal{FFT}\left(\sqrt{ J_\ell(\omega_{\ell k}) \Delta \omega} ~ y_{\ell k}^* \right).\\
    \end{split}
\end{equation}

In the fourth-order Runge-Kutta (RK4) method, the values of the noises are needed at intermediate half-time-steps of the simulation time-step $\Delta t$. Therefore we use $\delta t = \Delta t/2$ and $N_{\mathrm{step}} = 2N$, where $N$ is the number of simulation time-steps required.

\subsection{HOPS Parameters and Convergence Checks} \label{app:HOPS:conv}

\begin{table*}[htbp]
\centering
\begin{tabular}{|c|c|c|c|c|c|}
\hline
\multirow{2}{*}{Figure}
& \multirow{2}{*}{Fixed System Parameters}
& \multicolumn{2}{c|}{Variable System Parameters}
& \multicolumn{2}{c|}{\shortstack{HOPS Parameters:\\$N_{\mathrm{BCF}}=4, k=2$\\$N=20000$}} \\
\cline{3-4}\cline{5-6}
&
& Line Label
& Horizontal Axis
& $N_{\mathrm{trunc}}$
& $\Omega_c \Delta t$ \\
\hline

\multirow{3}{*}{Fig.~\ref{fig:U:thermo_T_10}}
& $T_h/\Omega_c=1.0, T_c/\Omega_c=0.5$
& $U=0$
& \multirow{3}{*}{$\Omega_h/\Omega_c\in[1.0,3.0]$}
& 2 & \multirow{3}{*}{0.02} \\
\cline{3-3}\cline{5-5}
& $\kappa_h/\kappa_c=1.0, \kappa_c/\Omega_c=0.05, \omega_{\mathrm{cut}}/\Omega_c=3.0$
& $U/\Omega_c=0.4,2.0,10.0$
&
& 8 & \\
\cline{3-3}\cline{5-5}
& $g/\Omega_c=0.5,\Omega_c = 1.0$
& $U\to\infty$
&
& 2 &  \\
\hline

\multirow{6}{*}{\shortstack{Fig.~\ref{fig:U:thermo_T_50}\\Fig.~\ref{app:fig:HOPS_empirical_analytic}}}
& \multirow{6}{*}{\shortstack{$T_h/\Omega_c=5.0, T_c/\Omega_c=0.5$ \\ $\kappa_h/\kappa_c=1.0,\kappa_c/\Omega_c=0.05, \omega_{\mathrm{cut}}/\Omega_c=3.0$ \\ $g/\Omega_c=0.5,\Omega_c = 1.0$}}
& $U=0$
& \multirow{1}{*}{$\Omega_h/\Omega_c\in[1.0,5.0]$}
& 6 & \multirow{6}{*}{0.02} \\
\cline{3-4}\cline{5-5}
&
& \multirow{3}{*}{$U/\Omega_c=0.4$}
& $\Omega_h/\Omega_c = 1.0, 1.2, 1.4$
& 12 &  \\
\cline{4-4}\cline{5-5}
&
&
& $\Omega_h/\Omega_c= 1.6, 1.8, 2.0$
& 10 &  \\
\cline{4-4}\cline{5-5}
&
&
& $\Omega_h/\Omega_c\in[2.2,5.0]$
& 8 & \\
\cline{3-4}\cline{5-5}
&
& $U/\Omega_c=2.0,10.0$
& \multirow{2}{*}{$\Omega_h/\Omega_c\in[1.0,5.0]$}
& 2 &  \\
\cline{3-3}\cline{5-5}
&
& $U\to\infty$
&
& 2 &  \\
\hline

\multirow{20}{*}{Fig.~\ref{fig:U:HOPS:non_reciprocal_U}(a)}
& \multirow{20}{*}{\shortstack{$T_c/\Omega_c=0.5$ \\ $\kappa_c/\Omega_c=0.05, \omega_{\mathrm{cut}}/\Omega_c=3.0$ \\ $g/\Omega_c=0.5,\Omega_h/\Omega_c=1.0,\Omega_c = 1.0$}}
& \multirow{5}{*}{\shortstack{$T_h/T_c=10.0$\\$ \kappa_h/\kappa_c=5.0$}}
& $U/\Omega_c = 0.4$
& 12 & 0.02 \\
\cline{4-4}\cline{5-6}
&
&
& $U/\Omega_c = 1.0$
& 10 & 0.02 \\
\cline{4-4}\cline{5-6}
&
&
& $U/\Omega_c = 2.0, 5.0, 10.0$
& 8 & 0.02 \\
\cline{4-4}\cline{5-6}
&
&
& $U/\Omega_c = 20.0$
& 8 & 0.01 \\
\cline{4-4}\cline{5-6}
&
&
& $U\to\infty$
& 2 & 0.02 \\
\cline{3-4}\cline{5-6}

&
& \multirow{5}{*}{\shortstack{$T_h/T_c=2.0$\\$ \kappa_h/\kappa_c=5.0$}}
& $U=0$
& 6 & \multirow{5}{*}{0.02} \\
\cline{4-4}\cline{5-5}
&
&
& $U/\Omega_c=0.04$
& 12 & \\
\cline{4-4}\cline{5-5}
&
&
& $U/\Omega_c=0.08$
& 10 &  \\
\cline{4-4}\cline{5-5}
&
&
& $U/\Omega_c\in[0.2, 20.0]$
& 8 &  \\
\cline{4-4}\cline{5-5}
&
&
& $U\to\infty$
& 2 & \\
\cline{3-4}\cline{5-6}

&
& \multirow{5}{*}{\shortstack{$T_h/T_c=10.0$\\$ \kappa_h/\kappa_c=2.0$}}
& $U=0$
& 6 & \multirow{5}{*}{0.02} \\
\cline{4-4}\cline{5-5}
&
&
& $U/\Omega_c=0.4$
& 12 &  \\
\cline{4-4}\cline{5-5}
&
&
& $U/\Omega_c=1.0$
& 10 &  \\
\cline{4-4}\cline{5-5}
&
&
& $U/\Omega_c\in[2.0,20.0]$
& 8 &  \\
\cline{4-4}\cline{5-5}
&
&
& $U\to\infty$
& 2 &  \\
\cline{3-4}\cline{5-6}

&
& \multirow{5}{*}{\shortstack{$T_h/T_c=2.0$\\$ \kappa_h/\kappa_c=2.0$}}
& $U=0$
& 6 & \multirow{5}{*}{0.02} \\
\cline{4-4}\cline{5-5}
&
&
& $U/\Omega_c=0.04$
& 12 & \\
\cline{4-4}\cline{5-5}
&
&
& $U/\Omega_c=0.08$
& 10 &  \\
\cline{4-4}\cline{5-5}
&
&
& $U/\Omega_c\in[0.2, 20.0]$
& 8 &  \\
\cline{4-4}\cline{5-5}
&
&
& $U\to\infty$
& 2 &  \\
\hline

\multirow{2}{*}{Fig.~\ref{fig:U:HOPS:non_reciprocal_U}(b)}
& $T_c/\Omega_c = 0.5,\kappa_c/\Omega_c=0.05, \omega_{\mathrm{cut}}/\Omega_c=3.0$
& $\kappa_h/\kappa_c=5.0$
& \multirow{2}{*}{$T_h/T_c\in[1.0,10.0]$}
& 2 & \multirow{2}{*}{0.02} \\
\cline{3-3}\cline{5-5}
& $g/\Omega_c=0.5, \Omega_h/\Omega_c=1.0, \Omega_c = 1.0, U\to\infty$
& $\kappa_h/\kappa_c=2.0$
&
& 2 &  \\
\hline

\multirow{6}{*}{Fig.~\ref{fig:U:omega_c}(a)}
& \multirow{6}{*}{\shortstack{$T_h/\Omega_c=5.0, T_c/\Omega_c=0.5$ \\ $\kappa_h/\kappa_c=1.0,\kappa_c/\Omega_c=0.05$ \\ $g/\Omega_c=0.5,\Omega_h/\Omega_c=1.0,\Omega_c = 1.0$}}
& $U/\Omega_c=0.4$
& \multirow{3}{*}{$\omega_{\mathrm{cut}}/\Omega_c \in [1.0, 3.0]$}
& 12 & \multirow{3}{*}{0.02} \\
\cline{3-3}\cline{5-5}
&
& $U/\Omega_c=2.0$
&
& 8 &  \\
\cline{3-3}\cline{5-5}
&
& $U\to\infty$
&
& 2 &  \\
\cline{3-4}\cline{5-6}
&
& $U/\Omega_c=0.4$
& \multirow{3}{*}{$\omega_{\mathrm{cut}}/\Omega_c \in [4.0, 6.0]$}
& 20 & \multirow{3}{*}{0.01} \\
\cline{3-3}\cline{5-5}
&
& $U/\Omega_c=2.0$
&
& 20 & \\
\cline{3-3}\cline{5-5}
&
& $U\to\infty$
&
& 2 & \\
\hline

\multirow{6}{*}{\shortstack{Fig.~\ref{fig:U:omega_c}(b)\\
Fig.~\ref{fig:U:HOPS:broadening}}}
& \multirow{6}{*}{\shortstack{$T_h/\Omega_c=5.0, T_c/\Omega_c=0.5$ \\ $\kappa_h/\kappa_c=1.0, \omega_{\mathrm{cut}}/\Omega_c=3.0$ \\ $g/\Omega_c=0.5,\Omega_h/\Omega_c=1.0,\Omega_c = 1.0$}}
& $U/\Omega_c=0.4$
& \multirow{3}{*}{$\kappa_c/\Omega_c \in [0.01,0.05]$}
& 12 &\multirow{3}{*}{0.02} \\
\cline{3-3}\cline{5-5}
&
& $U/\Omega_c=2.0$
&
& 8 & \\
\cline{3-3}\cline{5-5}
&
& $U\to\infty$
&
& 2 & \\
\cline{3-4}\cline{5-6}
&
& $U/\Omega_c=0.4$
& \multirow{3}{*}{$\kappa_c/\Omega_c \in [0.10, 0.50]$}
& 12 & \multirow{3}{*}{$\frac{\Omega_c}{10^3\kappa_c}$} \\
\cline{3-3}\cline{5-5}
&
& $U/\Omega_c=0.4,2.0$
&
& 12 & \\
\cline{3-3}\cline{5-5}
&
& $U\to\infty$
&
& 2 & \\
\hline

\end{tabular}
\caption{HOPS parameters used for figures in the main text.}
\label{table:HOPS:params}
\end{table*}

\begin{figure}[t]
    \centering
\includegraphics[width=0.95\linewidth]{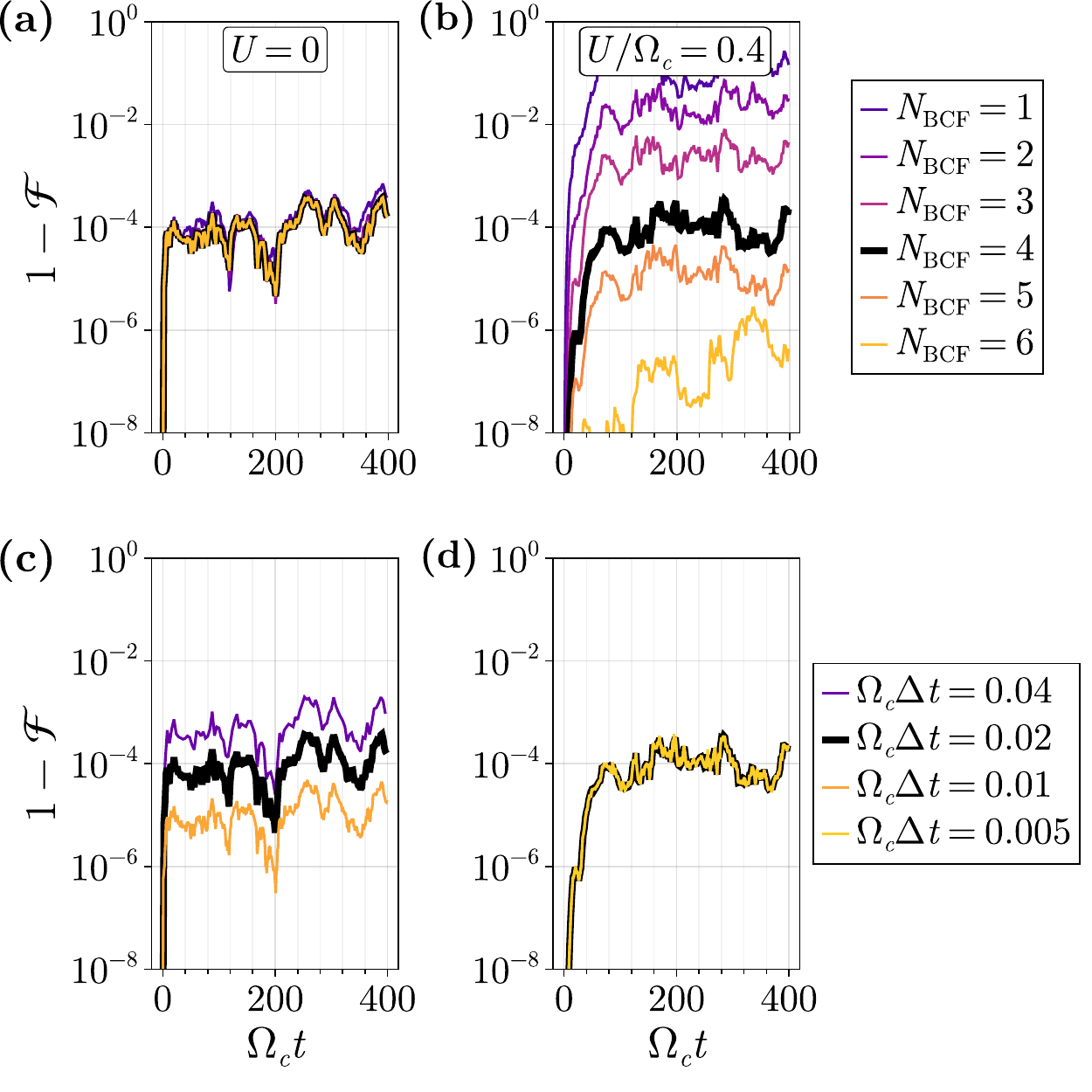}
    \caption{Infidelity $1-\mathcal{F}$ for different interaction strengths $U$. Fidelity is calculated against a reference trajectory using $N_{\mathrm{BCF}}=6, k =3,\Omega_c \Delta t=0.005$. Panels (a), (c): $U=0, N_{\mathrm{trunc}}^{\mathrm{ref}}=6$.
    Panels (b), (d): $U/\Omega_c=0.4, N_{\mathrm{trunc}}^{\mathrm{ref}}=12$. Panels (a), (b):  $N_{\mathrm{BCF}}$ is varied while $\Omega_c \Delta t=0.02$ is fixed. Panels (c), (d): $\Omega_c \Delta t$ is varied while $N_{\mathrm{BCF}}=4$ is fixed.
    System parameters: $ T_h/\Omega_c=1.0, T_c/\Omega_c=0.5, \omega_\mathrm{cut}/\Omega_c=3.0, \Omega_h/\Omega_c=1.0, g/\Omega_c=0.5, \kappa_{h}=\kappa_c=0.05 \Omega_c$. Black lines indicate the parameters used in Table~\ref{table:HOPS:params}. Reference parameters: $N_{\mathrm{BCF}}=6, k=3, \Omega_c \Delta t = 0.005, N = 80000$.} 
    \label{fig:exact:benchmark}
\end{figure}

The convergence of the HOPS method is checked by fixing the noise realisation and comparing against a reference trajectory $\ket{\tilde{\psi}_{\mathrm{ref}}^{(\vec{0})}(t)}$ with higher truncation parameters. The fidelity is calculated as a function of time $\mathcal{F}(t)=|\braket{\tilde{\psi}^{(\vec{0})}(t)|\tilde{\psi}_{\mathrm{ref}}^{(\vec{0})}(t)}|^2$. Convergence checks for $N_{\mathrm{trunc}}$ and hierarchy depth $k$ were also conducted and available upon request. 

We find that for $U=0$, the displacement method has reduced the accuracy to first-order in time-step $\Delta t$, as the displacement operation is applied only at the end of each time-step. This is the limiting factor in resulting fidelity. This could potentially be improved by modifying the equation of motion to continuously vary the displacement parameter. 

For $U>0$, the interaction picture retains the fourth-order accuracy in time-step. Here, the number of exponentials $N_{\mathrm{BCF}}$ limits the resulting fidelity. Table \ref{table:HOPS:params} lists the parameters used for HOPS in this work.

For certain parameter regimes in this work, HOPS was not used as it would require either too many Focks states $N_{\mathrm{trunc}}$ or hierarchy depth $k$. Modifications can be made to HOPS to reduce the required hierarchy depth for simulating highly excited, non-Markovian reservoirs  \cite{muller_nuhops_2025}.

\section{Perturbation Theory}\label{app:perturbation}
Using perturbation theory, we demonstrate the effects of on-site interaction $U$ on the spectrum of the system Hamiltonian in the rotating frame,
\begin{equation}
    H_S=g(a_h^\dagger a_c+ a_h a_c^\dagger) +\sum_{\ell=h,c} \frac{U}{2} n_\ell(n_\ell-1).
\end{equation}
\subsubsection{Weak Interaction Regime}
When $g>U$, we consider a perturbation on $H_0=g(a_h^\dagger a_c+ a_h a_c^\dagger)$ by $H_1=\sum_{\ell=h,c} \frac{U}{2} n_\ell(n_\ell-1)$ . $H_0$ can be cast into two uninteracting modes $+,-$ by using $a_\pm=\frac{a_h \pm a_c}{\sqrt{2}}$, resulting in $H_0 = g(a_+^\dagger a_+ - a_- ^\dagger a_-)$, with unperturbed eigenstates $\ket{n_+,m_-}$ where $n,m\in\mathbb{N}$. The first-order change to the energies are
\begin{equation}
    \Delta E_{n_+ m_-}^{(1)}=\Braket{n_+ m_-| H_1|n_+ m_-}.
\end{equation}
It is convenient to write $a_h=\frac{a_++a_-}{\sqrt{2}}$, $a_c=\frac{a_+-a_-}{\sqrt{2}}$. Then 
\begin{equation}
\begin{split}
    H_1 = & \frac{U}{2}\left(\frac{a_+^\dagger+a_-^\dagger}{\sqrt{2}}\right)\left(\frac{a_++a_-}{\sqrt{2}}\right) \left[\left(\frac{a_+^\dagger+a_-^\dagger}{\sqrt{2}}\right)\left(\frac{a_++a_-}{\sqrt{2}}\right)-1\right] \\
    & + \frac{U}{2}\left(\frac{a_+^\dagger-a_-^\dagger}{\sqrt{2}}\right)\left(\frac{a_+-a_-}{\sqrt{2}}\right) \left[\left(\frac{a_+^\dagger-a_-^\dagger}{\sqrt{2}}\right)\left(\frac{a_+-a_-}{\sqrt{2}}\right)-1\right].
\end{split}
\end{equation}

When we expand this, we may keep only the terms which do not change the number sectors of $+,-$ modes for computing the first order change to energies, 
\begin{equation}
    \Delta E_{n_+ m_-}^{(1)}=\frac{U}{4}\left[ n(n-1)+4nm + m(m-1)\right].
\end{equation} 
As a result, the transition frequencies between neighbouring eigenstates are given by
\begin{equation}
\begin{split}
    E_{n+1_+m_-}-E_{n_+m_-} & = g + \frac{U}{2}(n+2m),\\
    E_{n_+m+1_-}-E_{n_+m_-} & = -g + \frac{U}{2}(2n+m).\\
\end{split}
\end{equation}
\subsubsection{Strong Interaction Regime}
When $g<U$, we consider this to be a perturbation on $H_0=\sum_{\ell=h,c} \frac{U}{2} n_\ell(n_\ell-1)$ by $H_1=g(a_h^\dagger a_c+ a_h a_c^\dagger)$. The unperturbed eigenstates are of the form $\ket{n_h,m_c}$ corresponding to Fock states of each harmonic oscillator. There is no first-order change to the energies 
\begin{equation}
    \Delta E_{n_h m_c}^{(1)}=\Braket{n_h,m_c|H_1|n_h,m_c}=0,
\end{equation}
except for the degenerate pair of states $\ket{0_h,1_c},\ket{1_h,0_c}$ which become hybridized to $\ket{1_+0_-}=\frac{\ket{0_h, 1_c}+\ket{1_h,0_c}}{\sqrt{2}}$ and $\ket{0_+1_-}=\frac{\ket{0_h, 1_c}-\ket{1_h,0_c}}{\sqrt{2}}$ with energies $\pm g$.

As a result, the transition frequencies between neighbouring eigenstates are given by
\begin{equation}
\begin{split}
    E_{n+1_h m_c}-E_{n_h m_c} & = n U, \qquad n,m\geq1.\\
    E_{n_h m+1_c}-E_{n_h m_c} & = m U, \qquad n,m\geq1.\\
    E_{1_+0_-}-E_{0_h0_c} & = g,\\
    E_{0_+1_-}-E_{0_h0_c} & = -g,\\
    E_{1_h 1_c}-E_{1_+ 0_-} & = - g,\\
    E_{1_h 1_c}-E_{0_+ 1_-} & = g.
\end{split}
\end{equation}

\subsubsection{Transition Frequencies in Lab Frame}
When transforming from the rotating frame back to the lab frame, the eigenfrequencies $E$ as seen by each reservoir $\ell=h,c$ are modified by $E\to E+\Omega_\ell$. As a result, the system's transition frequencies $\Sigma$ in the presence of interactions are
\begin{equation}
   \Sigma= \begin{cases} 
      \{\Omega_\ell \pm g + r \frac{U}{2}~|~ r\in \mathbb{N}\} & U<g, \\
      \{\Omega_\ell \pm g\} \cup\{\Omega_\ell + r U~|~ r\in \mathbb{N}\} &  U>g .
   \end{cases}
\end{equation}
This matches the observed broadening of the momentum-resolved particle currents towards higher frequencies (Fig. \ref{fig:U:HOPS:broadening}). The above formulae may be extended easily for different on-site interaction strengths $U_h\ne U_c$.

\bibliographystyle{apsrev4-2}
\bibliography{references}

\end{document}